# Nonlinear energy harvesting system with multiple stability

Yanwei Han[a,*], Zijian Zhang[b]

[a] School of Civil Engineering and Architecture, Henan University and Science and Technology, Luoyang 47100, China;

[b] College of Astronautics, Nanjing University of Aeronautics and Astronautics, Nanjing 210016, China

* Corresponding author. E-mail address: yanweihan@haust.edu.cn

**Abstract:** The nonlinear energy harvesting systems of the forced vibration with an electron-mechanical coupling are widely used to capture ambient vibration energy and convert mechanical energy into electrical energy. However, the nonlinear response mechanism of the friction induced vibration (FIV) energy harvesting system with multiple stability and stick-slip motion is still unclear. In the current paper, a novel nonlinear energy harvesting model with multiple stability of single-, double- and triple-well potential is proposed based on V-shaped structure spring and the belt conveying system. The dynamic equations for the energy harvesting system with multiple stability and self-excited friction are established by using Euler-Lagrangian equations. Secondly, the nonlinear restoring force, friction force, and potential energy surfaces for static characteristics of the energy harvesting system are obtained to show the nonlinear varying stiffness, multiple equilibrium points, discontinuous behaviors and multiple well response. Then, the equilibrium surface of bifurcation sets of the autonomous system is given to show the third-order quasi zero stiffness (QZS3), fifth-order quasi zero stiffness (QZS5), double well (DW) and triple well (TW). The co-dimension bifurcation sets of the self-excited vibration system are analyzed and the corresponding phase portraits for the coexistent of multiple limit cycles are obtained. Furthermore, the analytical formula of amplitude frequency response of the approximated system are obtained by the complex harmonic method. The response amplitudes of charge, current, voltage and power of the forced electron-mechanical coupled vibration system for QZS3, QZS5, DW and TW are analyzed by using the numerically solution. Finally, a prototype of FIV energy harvesting system is manufactured and the experimental system is setup. The experimental work of static restoring force, damping force and electrical output are well agreeable with the numerical results, which testified the proposed FIV energy harvesting model.

## 1. Introduction

During the last decade, the renewable and green energy of hydro, biomass, solar, wind, geothermal, tidal, sea waves, mechanical vibration and human movement have been received extensive attention due to the exhaustion of fossil energy and serious environmental pollution [1]. Recently vibration energy harvesting technology, a new green energy method to obtain electrical energy from an ambient vibration source environment and replace the traditional battery and wired power, have been widely used in self-powered devices such as, actuators, wireless sensor, mobile electronic platform, wearable devices, headphones, bluetooth, medical implant, healthy monitoring and tire pressure monitoring [2-4]. Therefore, lots of scholars developed energy harvesting devices that harness and convert mechanical vibration energy into electrical energy by various approaches, such as the electromagnetic, electrostatic, piezoelectric, magnetostrictive and triboelectric methods [5-7]. Presently, the nonlinear design methods of the friction induced vibration, multiple stability, multiple degree-of-freedom system and stochastic excitation are widely used to design the energy harvesting system.

Nonlinear friction-induced vibrations, can usually cause severe damage and serious problems in machinery industry, are investigated by researchers in mathematical method, computational calculations and experimental testing. Warminski give the stability regions of the dry friction chatter for self-excited cutting model with two degree-of-freedom system and by using the multiple time scale time method and numerical calculation [8]. Warminski analyse the Hopf bifurcation, amplitude response and frequency locking for the nonlinear self-excited MEMS system with the external excitation and time delay by using multiple time scale method [9]. Tadokoro et al propose a friction-induced vibration energy harvester with a piezoelectric element by using analytical model, numerical analysis and experimental tests to demonstrate the feasibility of power generation [10]. Wei et al study stability diagram and chaotic region of stick-slip vibration for a new brake dynamic model and explore the mechanism and noise of brake pad vibration via the theoretical model and numerical analysis [11]. On the contrary, this kind of vibration can also be converted to electric energy by the energy harvesting technology. Recently, the methodologies of friction-introduced vibration are used to design the nonlinear energy harvesting system. Sinou et al study the Hopf bifurcation and the Routh-Hurwitz criterion for the friction-introduced two-degree-of-freedom system and find that the mode coupling instability depends on the structural damping [12]. Olejnik et al present discontinuous dynamical systems with Filippov-type nonlinearity and obtain the numerical estimations of the stick-slip transition by using method Henon's algorithm [13]. Wang et al investigate the friction-induced vibration energy harvesting system with piezoelectric element by using experimental and numerical simulation which offer a new way to design energy harvester [14]. Wang et al establish reciprocating friction-induced vibration energy harvesting system with piezoelectric elements to collect vibration energy and suppress friction-induced vibration noise by the numerical simulation [15]. Xiang et al establish the the friction-induced energy harvesting system with the piezoelectric cantilever beam and analyze the energy harvesting performance by the experimental test, finite element and numerical simulation models [16]. However, the hybrid combination of the nonlinear friction-introduced vibration and nonlinear multiple design to achieve high efficiency energy harvesting system has rarely been studied. Therefore, this paper proposed a now hybrid system integrated the characteristics of the above two nonlinearities of friction and multistability and achieved a higher power output efficiency.

For the drawback of the the linear energy harvesters of narrow frequency band and they lower electric power efficiency, the nonlinear energy harvesting system were designed to over come those difficulties from the linear system. Therefore, the nonlinear energy harvester with multistable potential function have been widely investigated by many researchers to improve the output power. Designed a vibrational tristable energy harvester for the purpose of improving energy harvesting performance under low-level ambient vibrations. tristable energy harvester system were numerically and experimentally verified: they have better energy harvesting performance when compared with their bistable counterparts under low-level harmonic or random excitation. Tekam et al analyze the periodic response using the KB averaging method and Melnikov criteria of the horseshoes chaos for the nonlinear triple energy harvester with fractional order viscoelastitiy [17]. Kim et al investigate the multi-stable energy harvester system by theoretic and experimental research and derive the analytical multi-stable parametric conditions on the basis of bifurcation analyses [18]. Zhou et al theoretically analyzes the influence of system parameters on the nonlinear dynamic responses of tristable energy harvesters and the by using the harmonic balance method [19]. Zhu et al investigates the harvesting advantage, wide frequency and high output voltage for a tristable energy harvester with the attractive forces between the beam tip's magnet and the fixed magnets [20]. Li et al present the probability analysis of the tri-stable energy harvesters with interwell vibration base on the Monte Carlo Simulation and verify the excellent energy harvesting performance by using theoretical and experimental methods [21]. Xu et al established the tristable energy harvester under Gaussian white noise excitation and predict the good performance of the mean output power by an approximate procedure [22]. Huang et al presents the analytical steady-state response displacement and output voltage for the tristable energy harvester and analyze the influence mechanism of system parameters on energy harvesting performance [23]. Margielewicz et al present a tristable energy harvesting system of vibrating mechanical devices with quasi-zero stiffness characteristic which can significantly improve ability in low frequency [24]. However, nonlinear elastic force of multiple potential in combination with friction-introduced energy harvesting system to achieve high efficiency has rarely been studied. Therefore, we proposed a new integrated system with zero stiffness and multiple well potential to achieve a higher production efficiency.

The remainder of the paper is outlined as follows. Section 2 present the mathematical equation of motion for the nonlinear harvesting system with the irrational elastic force of multiple stabilities, Stribeck friction of self-vibration and damping of quotient form. Section 3 gives the free vibration characteristic of restoring force of high order QZS and multiple stability characteristic, nonlinear friction and damping. The dynamic transition of the equilibrium bifurcation and the codimension bifurcation are analyzed in Section 4. Section 5 studies the output performance of the energy harvester for the different geometrical and physical parameters by numerical simulation method. In Section 6, an experimental prototype is fabricated vibration energy harvester to verify theoretical study and the numerical results. Finally, Section 7 concludes important remarks and give the potential application prospect.

## 2. The FIV nonlinear energy harvesting system

2.1 Modeling of the energy harvester

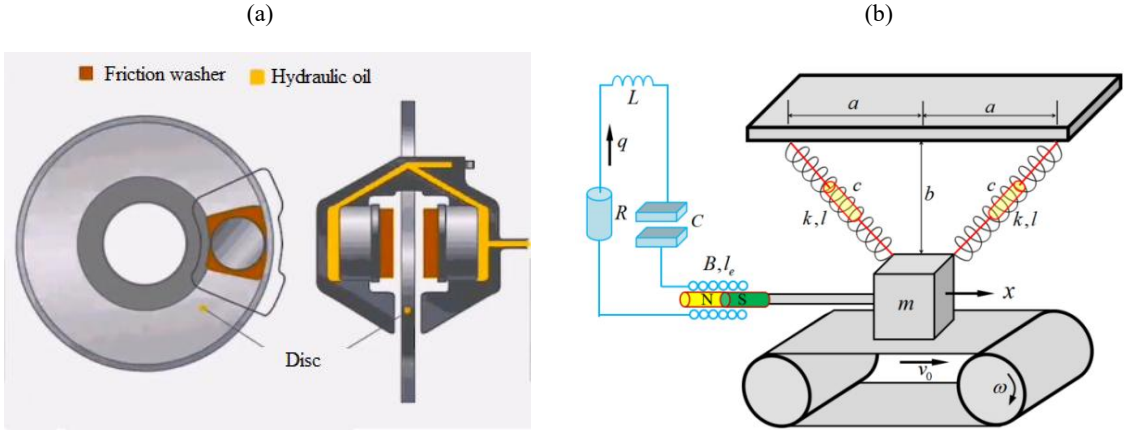

Fig. 1 Modeling of FIV system. (a)Physical model of the disc brake system [25] and (b)FIV energy harvesting model based on mass spring system under excited by conveyor belt (Color online).

Table 1. Parameters of energy harvesting system

| Parameter | Symbol | SI Unit | Dimension |
|---|---|---|---|
| Lumped mass | $m$ | kg (Kilogram) | M |
| Stiffness of springs | $k$ | N/m | $MT^{-2}$ |
| Damping coefficient | c | N·s/m | $MT^{-1}$ |
| Free length of oblique springs | $l_0$ | m (Meter) | L |
| Length of electromagnetic coil | $l_e$ | m (Meter) | L |
| Half distance of support | $a$ | m (Meter) | L |
| Distance of mass and support | $b$ | m (Meter) | L |
| Static friction coefficient | $\mu_s$ | N·s/m | $MT^{-1}$ |
| Dynamic friction coefficient | $\mu_m$ | N·s/m | $MT^{-1}$ |
| Velocity of belt | $v_m$ | m/s | $LT^{-1}$ |
| Inductance of coil | $L$ | H (Henry) | $L^2MT^{-2}I^{-2}$ |
| Capacitance of capacitor | $C$ | F (Faraday) | $L^{-2}M^{-1}T^4I^2$ |
| Resistance of coil | $R$ | Ω (Ohms) | $L^2MT^{-3}I^{-2}$ |
| Magnetic flux density | $B$ | T (Tesla) | $MT^{-2}I^{-1}$ |
| External frequency | $\omega_0$ | Rad/s | $T^{-1}$ |
| External force | $f_0$ | N | $MLT^{-1}$ |
| Time | $t$ | s (Second) | T |

As shown in Fig. 1(a), the disc brake system is most important in vehicle and mechanical engineer, which convert the kinetic energy of the vehicle into the thermal energy. The fluid from the master cylinder is forced in a caliper where it presses against a piston. The piston squeezes a pair of brake pads against the disk which is attached to the wheel, forcing to slow or stop. It is similar to the wheel rim friction between the bicycle brake and two rubber pads.

The nonlinear energy harvesting model as illustrated in Fig. 1(b), which consist of two part, the mass slider and the circuit. The point the mass sliders with mass $m$ (kg), is constrained by two viscous dampers $c$ (N·s/m), and a pair of two incline springs $k$ (N/m) in incline direction and is moving at the belt with speed $v_0$ (m/s). The electrical circuit with inductance $L$ (H), capacitor $C$ (F), resistance $R$ (Ω) and $B$ (T) is magnetic induction intensity. Beside, the two generalized coordinates of the displacement $x$ (m) in horizontal direction and is the electric charge $q$ (C) in the electric circuit. The geometrical parameter are horizontal relative distance $a$ (m) between two supporting pointsand vertical height $b$ (m) between board and mass. Moreover, $l_0$ (m) denotes free length of two oblique springs and $l_e$ (m) presents the length of electromagnetic coil. The nonlinear elastic force come form the geometrical nonlinearity of oblique spring structure. This geometrical nonlinear mechanism of the inclined spring are play an importance role in harvesting performance. Assume that the friction between the slider and belts is Stribeke type, this friction force lead to the self-exited response of the stick-slip and limit cycle.

2.2 Nonlinear differential equation of motion

Conventionally, the displacement $x$ and the charge $q$ are defined as the generalized coordinate for this nenergy harvester. Then the total kinetic energy $KE$(J) including the mechanical kinetic energy $KE_m$(J) and electrical kinetic energy $KE_e$(J) for the nonlinear FIV energy harvesting system are given in

$$KE = KE_m + KE_e \tag{1}$$

where $KE_m = 0.5m\dot{x}^2$ is the kinetic energy, $KE_e = 0.5L\dot{q}^2$ is the electrical energy, $\dot{x} = dx/dt = v$ (m/s) is the velocity in the horizontal direction of the mechanical system and $\dot{q} = dq/dt = i$ (A/s) is the current of the circuit system.

Subsequently, the nonlinear total potential energy $PE$(J) including the mechanical potential energy $PE_m$(J) of the linear elastic springs for the FIV nonlinear energy harvesting system referred to its lowest energy position and the electrical potential energy $PE_e$ (J) of the electric condenser can be written in following form

$$PE = PE_m + PE_e \tag{2}$$

where $PE_m = 0.5k(\sqrt{(x+a)^2} - l_0)^2 + 0.5k(\sqrt{(x-a)^2} - l_0)^2$ is the mechanical potential energy, and $PE_e = 0.5q^2/C$ is the electrical potential energy.

Thereafter, the nonlinear Rayleigh dissipation function $\Psi$ (J/s) of the mechanical vibration and electrical circuit are obtained as following

$$\Psi = \Psi_m + \Psi_d + \Psi_e + \Psi_q \tag{3}$$

herein $\Psi_m$, $\Psi_d$, $\Psi_e$ and $\Psi_q$, are the mechanical damping, electrical resistive, Stribeck friction and electromechanical coupled dissipation function respectively.

Furthermore, the Rayleigh function of nonlinear damping $\Psi_m$ (J/s) for then nonlinear energy harvesting system with geometrical configuration of V-shape structure is written as follows

$$\Psi_m = \frac{1}{2}c\frac{(x+a)^2}{(x+a)^2+b^2}\dot{x}^2 + \frac{1}{2}c\frac{(x-a)^2}{(x-a)^2+b^2}\dot{x}^2 \tag{4}$$

where $a$ and $b$ are the nonlinear geometrical structure parameters. Although the $c$ is the linear damping ration, in which is proportional to velocity. The resulting horizontal damping force on the mass have the strongly nonlinearity of the quotient term (*/*) due to the geometrical configuration.

Next, the Rayleigh function $\Psi_d$ (J/s) of the nonlinear Stribeck type is used to defined the friction between the lumped mass and belt as

$$\Psi_d = \int_0^{\dot{x}_r} f_d(\dot{x}_r)d\dot{x}_r = \begin{cases} \mu_s \text{sgn}(\dot{x}_r)\dot{x}_r - 0.5D_1\dot{x}_r^2 + 0.25D_3\dot{x}_r^4, & \dot{x}_r \neq 0 \\ [-\mu_s, \mu_s], & \dot{x}_r = 0 \end{cases} \tag{5}$$

where $D_1 = 3(\mu_s - \mu_m)/(2v_m)$ and $D_3 = (\mu_s - \mu_m)/(2v_m^3)$ are the stick-slip coefficients of kinetic friction obtained from experimental test, $\dot{x}_r = v_r = \dot{x} - v_0$ is the relative velocity, $\mu_s$ (N·s/m) is the static friction coefficient, $v_m$ is the velocity corresponding to the minimum dynamic friction coefficient $\mu_m$ (N·s/m). This friction can exhibits the nonlinearities of the nonlinear discontinuous of signum function and the nonlinear cubic term of Duffing type [26].

Similarity, the Rayleigh function of the electromagnetic dissipation function $\Psi_e$ (J/s) for the circuit part of the energy harvesting system is given by

$$\Psi_e = 0.5Ri^2 \tag{6}$$

where $R$ (Ω) is the resistance and $i$ (A) denotes the current.

The Rayleigh dissipation function $\Psi_q$ (J/s) of the electromechanical coupling has a functional dependence on the velocity and the current and is defined as

$$\Psi_q = -Bl_e \dot{x}\dot{q} \tag{7}$$

where $B$(T) is the magnetic flux density, $l_e$ (m) is the length of the coil, $Bl_e$(Tm) is the electrical damping coefficient which using to generates the power through magnetic mass oscillating of the variation of the magnetic field in a coil leading to the electrical current.

Furthermore, the generalized external force $Q_f$ (N) that acting on lumped mass block is assumed to the periodic exited harmonic force as follow

$$Q_f = f_0 \sin \omega_0 t \tag{8}$$

where $f_0$ (m/s²) and $\omega_0$ are the amplitude and frequency of the external excitation respectively.

Make using the Euler-Lagrange-Maxwell's equations, the equations of motion for the nonlinear energy harvesting system are give by

$$\begin{cases} \dfrac{d}{dt}\left(\dfrac{\partial \Pi}{\partial \dot{x}}\right) - \dfrac{\partial \Pi}{\partial x} + \dfrac{\partial \Psi}{\partial \dot{x}} = Q_f \\ \dfrac{d}{dt}\left(\dfrac{\partial \Pi}{\partial \dot{q}}\right) - \dfrac{\partial \Pi}{\partial q} + \dfrac{\partial \Psi}{\partial \dot{q}} = 0 \end{cases} \tag{9}$$

where Lagrange function is $\Pi = KE - PE$.

Finally, the motion differential equations of the FIV energy harvesting system with self-excited oscillation and stick-slip motion are obtained

$$\begin{cases} m\ddot{x} - Bl_e \dot{q} + f_m + f_d + f_s = f_0 \sin \omega_0 t \\ L\ddot{q} - Bl_e \dot{x} + R\dot{q} + \dfrac{1}{C}q = 0 \end{cases} \tag{10}$$

where Eq. (10) also obtained by applying Newton's law for mechanical system and Kirchhoff's for electrical circuits. The first equation defines mechanical response of the mass-spring system, while the second equation defines the electrical response of the circuit system.

To make study convenient, a four dimensional first-order equations of FIV nonlinear energy harvesting system are obtained as follows

$$\begin{cases} \dot{x} = v \\ m\dot{v} = Bl_e \dot{q}_s - f_m - f_d - f + f_0 \sin \omega_0 t \\ \dot{q} = i \\ L\dot{i} = Bl_e \dot{x} - R\dot{q} - q/C \end{cases} \tag{11}$$

where $v$ and $i$ are the velocity and the current respectively.

Where, the nonlinear damping of dampers is obtained by

$$f_m = c\left(\dfrac{(x+a)^2}{(x+a)^2 + b^2} + \dfrac{(x-a)^2}{(x-a)^2 + b^2}\right)\dot{x} \tag{12}$$

where damping coefficient $c$ is constant and damping force $c\dot{x}$ is linear and is proportion to the velocity. However, the result damping force $f_m$ is strongly nonlinearity with the quotient term (∗/∗) and quadratic term (∗²) because of the geometric configuration.

Here, the nonlinear elastic force of the V-shape supporting springs that acting on the mass is the nonlinear function as follows

$$f_s = k(x+a)\left(1 - \frac{l_0}{\sqrt{(x+a)^2 + b^2}}\right) + k(x-a)\left(1 - \frac{l_0}{\sqrt{(x-a)^2 + b^2}}\right) \tag{13}$$

herein the stiffness $k$ of the springs is linear, in which is proportional to displacement. Nonetheless the resulting horizontal force on the mass have the strongly nonlinearities with the quotient term ($*/*$) and irrational term ($\sqrt{*}$) due to the geometrical configuration.

Herein, the nonlinear Stribeck friction expression $f_d$ (N) between the mass and belt motion is defined as following formula

$$f_d = \begin{cases} \mu_s \, \text{sgn}(v_r) - D_1 v_r + D_3 v_r^3, & v_r \neq 0 \\ [-\mu_s, \mu_s], & v_r = 0 \end{cases} \tag{14}$$

where sign($*$) is the signum function. There are existence of negative-slop relationship between the friction force and the slider velocity. The friction force relative to the velocity of the slider is highly nonlinear near the zero velocity, at which the slop of the tangent line is infinity. When the slider slips $v_r \neq 0$, then fiction force $f_d \neq 0$. If the sticks $v_r = 0$, the friction force $f_d = 0$. If damping force $f_d$ lager than the break-away critical force. The stick-slip motion is thus a success of transitions between slip mode and stick mode. For example, the resonance vibration of the violin string is exited by stick-slip motion of the bow. The scrape of the chalk on the blackboard is another example.

2.3 Dimensionless equations of motion

To simplify the study of the governing equations of motion of the system (10), whose dimensionless process of the variables, parameters and time are chosen such as following

$$\begin{cases} X = \frac{x}{l_0}, X' = \frac{\dot{x}}{l_0 \omega_n}, Q = \frac{q}{q_0}, Q' = \frac{\dot{q}}{q_0 \omega_n}, q_0 = l_0 \sqrt{kC}, \\ \omega_n^2 = \frac{k}{m}, T = \omega_n t, V = \frac{v}{l_0}, V_r = \frac{v_r}{l_0}, V_0 = \frac{v_0}{l_0}, \Omega_0 = \frac{\omega_0}{\omega_n}, \\ \alpha = \frac{a}{l_0}, \beta = \frac{b}{l_0}, \gamma = \frac{Lq_0^2}{ml_0^2}, \theta = Bl_e \sqrt{\frac{C}{m}}, \xi_x = \frac{c}{2\sqrt{mk}}, \\ \xi_q = \frac{CR}{\sqrt{k/m}}, \mu = \frac{\mu_s}{kl_0}, \xi = \frac{D_1}{2\sqrt{mk}}, \eta = \frac{D_3}{\sqrt{mk}}, F_0 = \frac{f_0}{kl_0} \end{cases} \tag{15}$$

where $X$ is the dimensionless displacement. $X' = V$ is the dimensionless velocity. $T$ is the dimensionless time, $Q$ is the dimensionless constant charge. $\omega_n$ is the circular frequency depend on only on the mass and stiffness of the system. $\alpha$ and $\beta$ are dimensionless horizontal and vertical geometrical parameters respectively. $\gamma$ is the inertia ratio. $\xi_x$ is the mechanical damping ratio of viscous damper. $\xi_q$ and $\theta$ are the electrical resistance ratio and electromechanical coupling coefficient respectively. $\mu$, $\xi$ and $\eta$ are Stribeck friction ratios. $V_r$ and $V_0$ dimensionless relative and belt velocity. $\Delta$ is the dimensionless gravitational force.

Substituting the dimensionless transformation of Eq. (15) into the dimensional system (10), the dimensionless dynamic equations of FIV system are expressed as

$$\begin{cases} X'' + \theta Q' + F_d + F_m + F_s = F_0 \sin \Omega_0 T \\ \gamma Q'' + \xi_q Q' + \theta X' + Q = 0 \end{cases} \tag{16}$$

where the first equation is the mechanical motion equation of the spring-mass model and the second equation is the electrical circuit equation of RCL model. The FIV energy harvesting system (16) is a complicated nonlinear system due to spring restoring with quotient ($*/*$) and irrational ($\sqrt{*}$) nonlinearities and nonlinear friction force with Van der Pol type damping, which are dependent on the non-dimensional system parameters $\alpha$, $\beta$, $\mu$, $\xi$ and $\eta$. It is difficult to obtain the analytical expressions of response amplitude of the dimensionless displacement $X$ and charge $Q$ because of natural nonlinearity of the ordinary differential equation (ODE) of motion of Eq. (16). The performance of complex periodic solution dynamic bifurcation characteristics of the FIV system (16) are solved numerically by the numerical simulation.

Then the FIV energy harvesting system (16) can be described in four dimensional state space form as a first-order equations

$$\begin{cases} X' = V \\ V' = \theta Q' - F_m - F_d - F_s + F_0 \sin\Omega_0 T \\ Q' = I \\ \gamma I' = \theta X' - \xi_q Q' - Q \end{cases} \quad (17)$$

where $V$ is the nondimensional velocity and $I$ is the dimensionless current.

## 3. Static mechanical analysis

3.1 Nonlinear damping force

By the dimensionless definition of $F_m/kl_s$, the nonlinear damping $f_m$ Eq. (12) can be written as the non-dimensional form, that is

$$F_m = \xi_x \left( \frac{(X+\alpha)^2}{(X+\alpha)^2 + \beta^2} + \frac{(X-\alpha)^2}{(X-\alpha)^2 + \beta^2} \right) X' \quad (18)$$

As illustrated in Fig. 2, the nonlinear damping force $F_m$ are plotted to show parameter dependence of non dimensional geometrical ratio $\beta$ and displacement $X$ for different values of $\alpha = 0.0, 0.5, 1.0$ and $1.5$ respectively. It is found that the value of maximum $2$ and minimum $-2$.

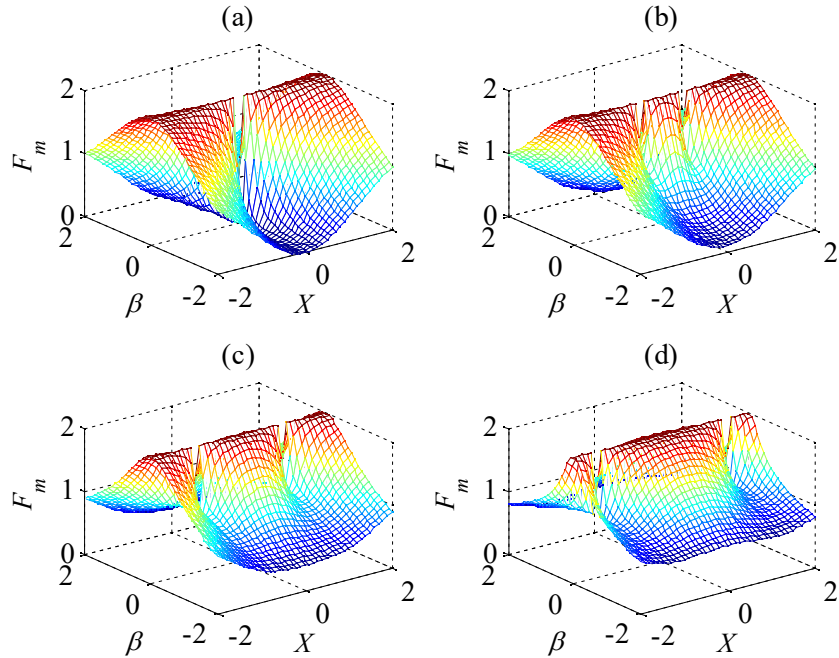

Fig. 2 Variation of damping force $F_m$ with displacement $X$ and ratio $\beta$ for $\xi_x = 1$ and $X' = 1$.
(a) $\alpha = 0$, (b) $\alpha = 0.5$, (c) $\alpha = 1.0$, (d) $\alpha = 1.5$ (Color online).

3.2 Nonlinear friction analysis

As presented in Fig. 3(a), the nonlinear friction force $f_d$ of Eq. (14) surfaces are plotted to shown the relationship of the force $f_d$ respect to parameters $v_r$ and $v_m$. The stribeck friction $f_d$ between the mass and the belt may lead to chatter vibration in the proposed friction-introduced vibration energy harvesting model. The nonlinear nature characteristics also give rise to the periodic stick-slip motion, which corresponding to the stable limit cycle.

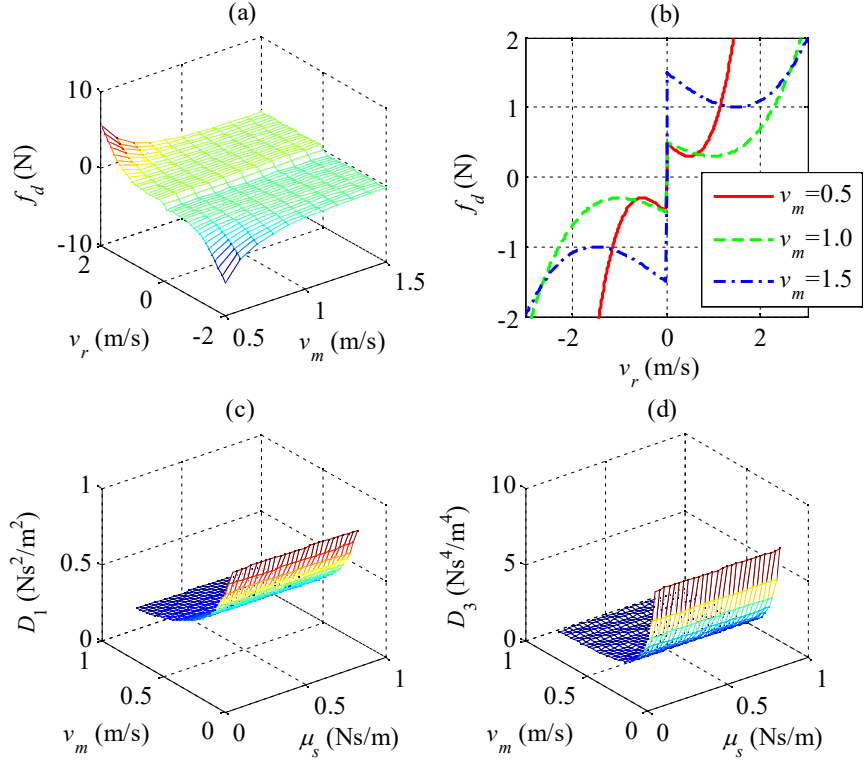

Fig. 3 Stribeck friction force. (a)Force surface. (b)Force curves for $\mu_s=1.5$, $\mu_m=1.0$.
(c)Parameters $D_1$ and (d) $D_3$ for $\mu_m=1.0$ (Color online).

As can be seen in Fig. 3(b), the nonlinear friction force $f_d$ is a three piece of smooth nonlinear curves. Four different regimes are identified: (i) In regime I, there is the static friction for $v_r = 0$; (ii) in regime II, sliding without building of the lubricant situation for $v_{m0} < v_r < v_{m1}$; (iii) in regime III, there is exist the partial lubrication for $v_{m2} < v_r < v_{m3}$; and (iv) in regime IV, the friction force of shear resistance is increase linearly with velocity for $v_r > v_{m3}$. In Fig. 2(c), the $D_1$ is increase with the $v_m$ increase. Fig. 3(d), the $D_3$ is increase with the $v_m$ increase.

The friction force $F_d$ with both typical nonsmooth and cubic nonlinearity is determined by relative velocity $V_r$ between mass $m$ and belt

$$F_d = \mu \, \text{sgn}(V_r) - \xi V_r + \eta V_r^3 \tag{19}$$

where $V_r = X' - V_0$ is nondimesional relative velocity.

For the first equation of system (16), letting $X'' = F_m = \theta = 0$, the total force consisting of two parts of damping force $F_d$ and restoring force $F_s$, yield to

$$F_d(-V_0) + F_s(X_s) = 0 \tag{20}$$

Solve from the equation (20), and then get the equilibrium position of mass

$$X_s = F_1^{-1}(-F_d(-V_0)) \tag{21}$$

Introducing the new variable of relative displacement

$$X_r = X - X_s \tag{22}$$

Submitting Eq. (22) into Eq. (16), the moving equation for the mechanical part of energy harvesting system (16) is obtained as follow:

$$X_r'' + X_s'' + F_d(X_r' + X_s' - V_0) + F_s(X_r + X_s) = 0 \tag{23}$$

For $X_s'' = X_s' = 0$ due to the equilibrium position $X_s$ is a constant value, so

$$X_r'' + (F_d(X_r' - V_0) + F_d(-V_0)) + (F_s(X_r + X_s) - F_d(-V_0)) = 0 \tag{24}$$

Then relative friction force is define as

$$F_{dr} = F_d(V_r) + F_d(-V_0) \tag{25}$$

where $V_r = X' - V_0$ is the relative velocity. And leading to the equation of motion is

$$X_r'' + F_{dr} + F_s(X_r + X_s) - F_d(-V_0) = 0 \tag{26}$$

The equation is a complex dynamical system that cannot be solved directly due to mathematical difficulties of the Van der Pol damping term and radical force term. In Fig. 4, the damping friction speed curves with negative slop of are plotted for different values of belt velocity $V_m = 0.5, 1.0, 1.5$. The sliding and stick occur in succession associated with friction coefficient decreases with rubbing speed lead to unstable and caused friction-induced vibration of self-excited motion. It is difficult to linearize a Stribeck friction law near the $V_r = 0$ that the linearization cannot be expected to predict behavior of chattering, sticking and limit cycling which caused by stick-slip friction.

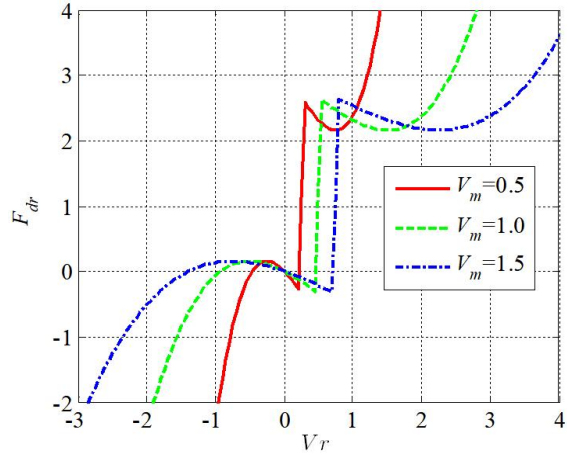

Fig. 4 Nonlinear friction velocity curves of Stricbeck type for Eq. (25) (Color online).

3.3 Nonlinear elastic restoring force with multiple stability

In order to obtain effect of system parameters on elastic force, the nonlinear restoring forces of the springs in *X*-direction is obtained an follows:

$$F_s = (X + \alpha)\left(1 - \frac{1}{\sqrt{(X + \alpha)^2 + \beta^2}}\right) + (X - \alpha)\left(1 - \frac{1}{\sqrt{(X - \alpha)^2 + \beta^2}}\right) \tag{27}$$

The proposed FIV energy system (16) allows us to manipulate monostable, bistable and tristable potential well via adjust two geometrical parameters $\alpha$ and $\beta$.

As illustrated in Fig. 5, when changing the dimensionless distance $\beta$ form $-2$ to $2$, the nonlinear restoring force $F_s$ of Eq. (23) surfaces demonstrate the varying process of monostable, bistable and tristable force at various non-dimensional distance parameter $\alpha = 0, 0.5, 1.0$ and $1.5$. It is observed that two peaks of Figs. 5(b-c) and one peak in Fig. 5(a).

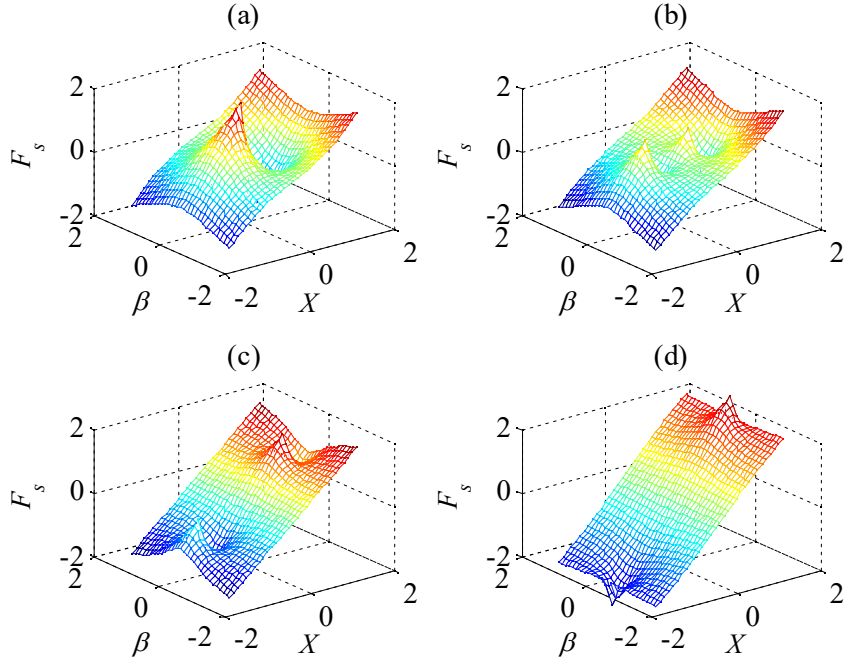

Fig. 5 Nonlinear restoring force $F_s$ surfaces. (a) $\alpha = 0$, (b) $\alpha = 0.5$, (c) $\alpha = 1.0$, (d) $\alpha = 1.5$ (Color online).

The curved surfaces of nonlinear restoring force plotted in Fig. 6 shows the how elastic force $F_s$ depends on geometrical ratio $\beta$ for different value 0, 0.5, 1.0, 1.5. It can be seen that discontinuous force of Fig. 6(a) with monostable, bistable and tristable potential well for $\alpha$=1.0, 0.5 and 0.0. Figs. 6 (b-d) demonstrates that the mutiple stability with smooth characteristic for non-dimensional height geometrical parameter $\beta = 0.5$, 1.0 and 1.5 respectively.

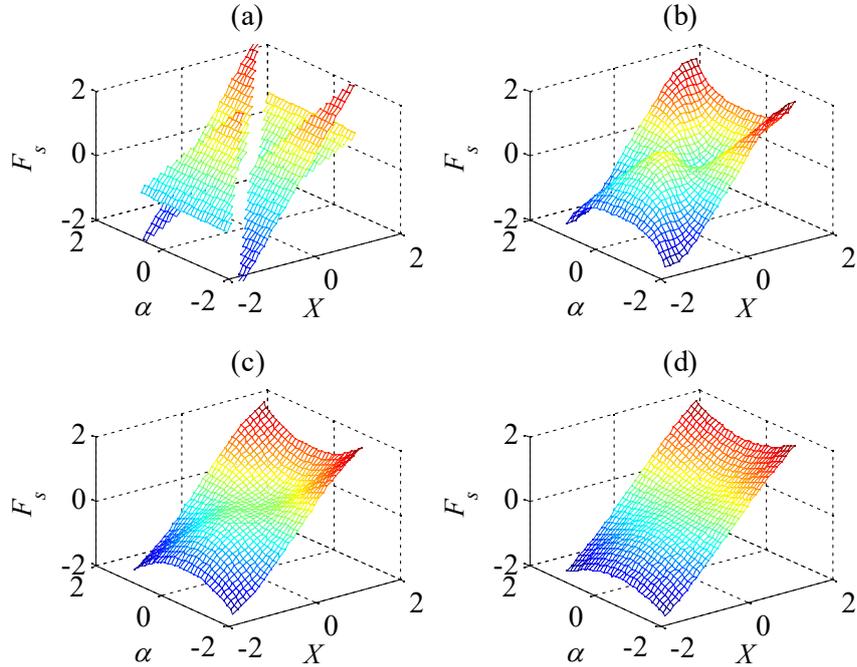

Fig. 6 Nonlinear surfaces of restoring force $F_s$. (a) $\beta = 0$, (b) $\beta = 0.5$, (c) $\beta = 1.0$, (d) $\beta = 1.5$ (Color online).

3.4 Potential energy with multiple stability

Integrate the Eq. (27) in the region $[X_0, X]$, the nonlinear dimensionless elastic potential energy (*PEN*) can be displayed as following form

$$PEN = \int_{X_0}^{X} F_s dX = 0.5\left(\sqrt{(X+\alpha)^2 + \beta^2} - 1\right)^2 + 0.5\left(\sqrt{(X-\alpha)^2 + \beta^2} - 1\right)^2 \qquad (28)$$

The nonlinear potential energy $PEN$ of Eq. (28) profiles are illustrated in Fig. 7 for different geometrical value $\alpha$. We vary the parameter $\alpha = 0.0, 0.5, 1.0$ and $1.5$ corresponding to Fig. 7(a-d) respectively. Fig. 7(a) shows that the $PEN$ curve is like a Mexican straw hat of one peak in the middle with maximum energy at the top of the hat and minimum energy at the bottom of the hat. Figs. 7(b-d) illustrated Mexican surface of two peaks with multiple well nature. It is found that the barrier separates multiple wells of the potential experienced by the apex as a function of geometrical parameter $\beta$. Therefore, the saddle and center points corresponding to extrema of the potential energy surface. The saddle points corresponding to maximum potential energy $PEN$ surface, at which the eigenvalues have positive and negative real parts. The center points corresponding to the minimum of the $PEN$ surface at which the eigenvalues are pure imaginary pairs or zero real parts.

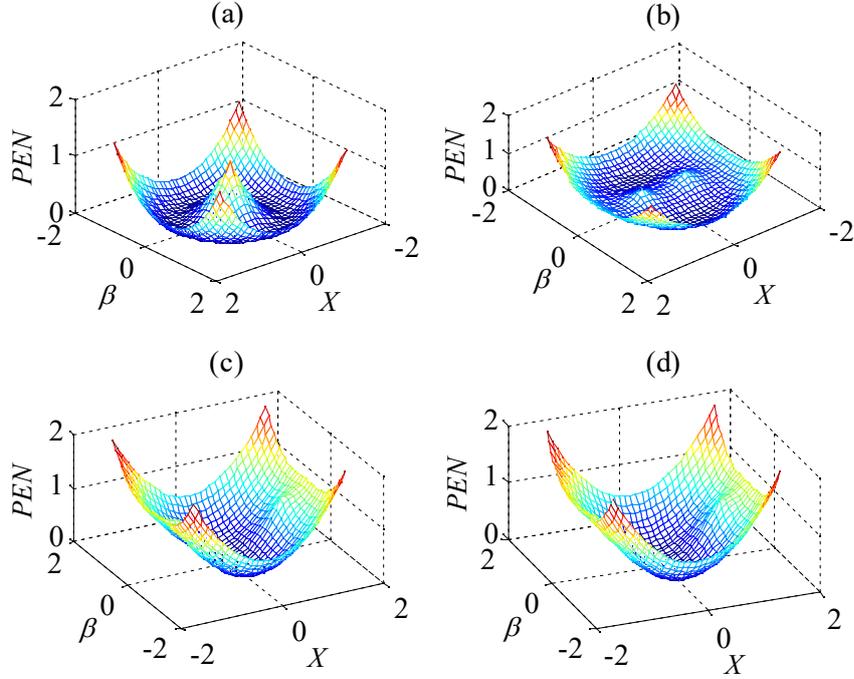

Fig. 7 Potential energy $PEN$ surfaces of Eq. (28) for different value of geometrical parameter $\alpha$.
(a) $\alpha = 0.0$, (b) $\alpha = 0.5$, (c) $\alpha = 1.0$, (d) $\alpha = 1.5$ (Color online).

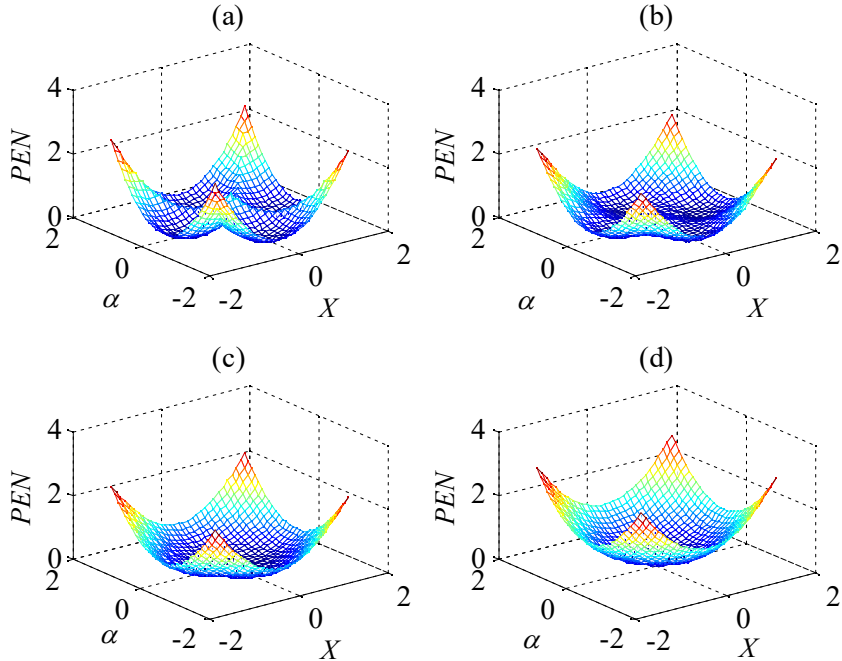

Fig. 8 Nonlinear potential energy surfaces for $PEN$ of Eq. (28) for different value of geometrical parameter $\alpha$.
(a) $\beta = 0.0$, (b) $\beta = 0.5$, (c) $\beta = 1.0$, (d) $\beta = 1.5$ (Color online).

The nonlinear potential energy function *PEN* of Eq. (28) surfaces are plotted in Fig. 8 for different geometrical value $\beta$. We vary the parameter $\beta = 0.0, 0.5, 1.0$ and $1.5$ corresponding to Fig. 8(a-d) respectively. Fig. 8(a) shows that the *PEN* curve is like a Mexican straw hat of one peak in the middle and backbone in radial direction. Figs. 8(b-d) illustrated Mexican surface of two peaks with multiple well nature.

## 4. Bifurcation analysis and multiple stick-slipe motion

### 4.1 Multiple stability and multiple well

The mechanical part of the FIV nonlinear energy harvesting system (16) is obtained as follows without friction $F_m = 0$, $F_d = 0$ and $\theta = 0$, and we have the free vibration equation in the form

$$X'' + F_s = 0 \tag{29}$$

where $F_s$ is the nonlinear restoring force of Eq. (13).

The equilibrium surface set $E$ of the free vibration system (29) for the nonlinear energy harvesting system is defined as

$$E = \{(X, \alpha, \beta) \mid F_s = 0\} \tag{30}$$

The bifurcation set $E$ is plotted in Fig. 9(a) in three dimensional parametric space $(X, \alpha, \beta)$. It is found that the equilibrium set $E$ had a rugged surface with the ridges and gullies. In Fig. 9(a) the bifurcated set of the original system of Eq. (29) of equilibrium surface $E$ in $(X, \alpha, \beta)$ are obtained.

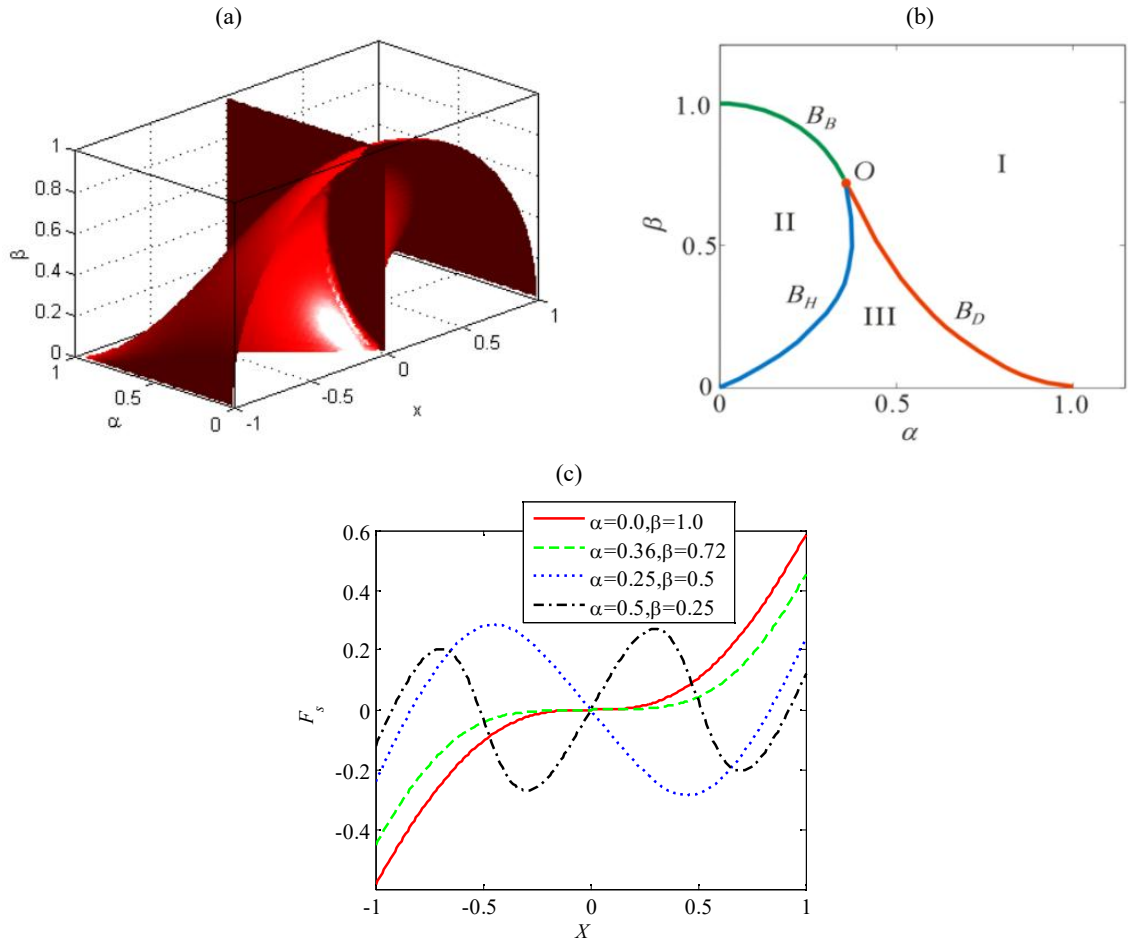

Fig. 9 Equilibrium bifurcation and nonlinear restoring force. (a)Equilibrium surface $E$ in parameter space $(X, \alpha, \beta)$ for Eq. (30). (b)Bifurcation curves on geometrical parameter $(\alpha, \beta)$ plane. The sup-pitchfork set $B_B$, the sub-pitchfork set $B_H$ and the saddle-node set $B_D$ on two dimensional parameter plane $(\alpha, \beta)$ for Eq. (31) (Color online). (c) Nonlinear forces of QZS3 —, QZS5—, DW··· and TW·-·-.

By the definition of the bifurcated parameter set $B_0 = B_B \cup B_H \cup B_D$ on $(\alpha, \beta)$ plane, the supercritical pitchfork bifurcation set $B_B$, the hysteresis set $B_H$ for subcritical pitchfork bifurcation and the double limit set $B_D$ for the saddle node bifurcation respectively, as following

$$\begin{cases} B_B = \{(\alpha, \beta) \mid F_X = 0, \partial F_X / \partial X = 0, \beta_0 < \beta < 1\} \\ B_H = \{(\alpha, \beta) \mid F_X = 0, \partial F_X / \partial X = 0, 0 < \beta < \beta_0\} \\ B_D = \{(\alpha, \beta) \mid F_X = 0, \partial F_X / \partial X = 0, \alpha_0 < \alpha < 1\} \end{cases} \quad (31)$$

herein $(\alpha_0, \beta_0) = (4\sqrt{5}/25, 8\sqrt{5}/25)$ corresponding to the higher codimension bifurcation point $O$.

In Fig. 9(b), bifurcation set branches $B_B$ is the super pitchfork bifurcation. $B_H$ is the sub pitchfork bifurcation, saddle-node bifurcation set $B_D$. The regions I, II, III corresponding the mono-stable, bi-stable and tri-stable state. We note that $\alpha_0$ and $\beta_0$ are bifurcation values because they separate values of $(\alpha, \beta)$ for which the phase portraits various qualitatively.

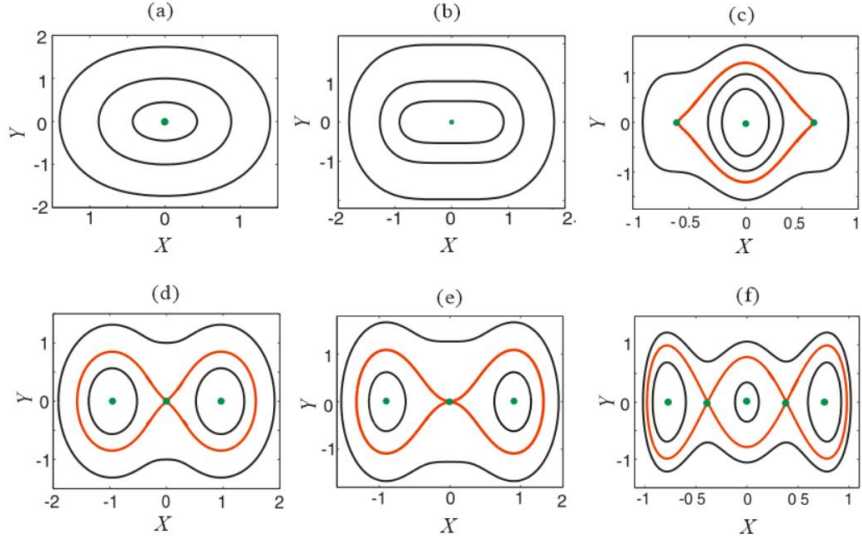

Fig. 10 Phase portraits of mechanical system (29) corresponding to the bifurcation set $B_0$ of Eq. (31) (a)In region I and set $B_B$, (b)at point $O$, (c)on set $B_D$, (d)on region II, (e)on set $B_H$, (f)on region III (Color online).

As shown in Fig. 10, phase portraits corresponding to the bifurcation set of Eq. (31) of Fig. 10(b) of the system (29) for the control parameter $(\alpha, \beta)$ plane are plotted with help of Hamilton energy function $H = KE + KP$. As depicted in Fig. 10(a), the quasi-zero stiffness with third order (QZS3) of single well (SW) on region I and third-order zero stiffness of on bifurcation set $B_B$ with geometrical parameter values of $(\alpha_1, \beta_1) = (0.0, 1.0)$. In the same way, Fig. 10(b) shows quasi-zero stiffness with five-order zero stiffness (QZS5) of single well at point $O$ with geometrical parameter values of $(\alpha_2, \beta_2) = (0.36, 0.72)$. Therefore, the phase portraits of single well potential energy are given in Fig. 10(a-b).

In Fig. 10(c), The phase trajectory of single well with mono-table equilibrium and heteroclinic orbit which connect two saddle-center point on double limit set $B_D$. As illustrated in Fig. 10(d), the bi-stable phase portrait of double well (DW) on region II with geometrical parameter values of $(\alpha_3, \beta_3) = (0.25, 0.5)$. As shown in Fig. 10(e), double well with bi-table equilibrium on hysteresis bifurcation set $B_H$. For double well cases, the phase portraits of with the homoclinic orbits which connect the saddle points are shown in Fig. 10(d-e). In Fig. 10(f), tri-stable trajectories of triple well (TW) equilibrium on region III with geometrical parameter values of $(\alpha_4, \beta_4) = (0.5, 0.25)$. The TW trajectories with the homonclinic and heteroclinic orbits is depicted in Fig. 10(f).

The nonlinear elastic force $F_s$ with quotient ($*/*$) and irrational ($\sqrt{*}$) nonlinearity is expressed as a polynomial with linear, cubic and quintic terms, that is

$$F_s = A_1 x + A_3 x^3 + A_5 x^5 + \cdots \quad (32)$$

and $A_1 = 2 - \dfrac{2}{(\alpha^2 + \beta^2)^{\frac{1}{2}}} + \dfrac{2\alpha^2}{(\alpha^2 + \beta^2)^{\frac{3}{2}}}$,

$$A_3 = \frac{1}{\left(\alpha^2+\beta^2\right)^{\frac{3}{2}}} - \frac{6\alpha^2}{\left(\alpha^2+\beta^2\right)^{\frac{5}{2}}} + \frac{5\alpha^4}{\left(\alpha^2+\beta^2\right)^{\frac{7}{2}}},$$

$$A_3 = -\frac{3}{4}\frac{1}{\left(\alpha^2+\beta^2\right)^{\frac{5}{2}}} + \frac{15}{4}\frac{\alpha^2}{\left(\alpha^2+\beta^2\right)^{\frac{7}{2}}} - \frac{105}{4}\frac{\alpha^4}{\left(\alpha^2+\beta^2\right)^{\frac{9}{2}}} + \frac{63}{4}\frac{\alpha^6}{\left(\alpha^2+\beta^2\right)^{\frac{11}{2}}}.$$

are the coefficients of the first, third and fifth order terms of the system (32) respectively.

Then the approximated system of the original mechanical system (29) with fifth order truncation of Taylor series, the fifth-order Duffing system can be written as following form

$$X'' + A_1 X + A_3 X^3 + A_5 X^5 = 0 \tag{33}$$

It is should be note that the polynomial nonlinear system (33) is topologically equivalent to the original system (29) without losing the global dynamics. This dynamical system has a form of the generalized Duffing oscillator with quintic nonlinearity and mono-stable, bi-stable and tri-stable characteristic.

As shown in Fig. 11, the bifurcation set and multi-steady state phase diagram characteristics of the polynomial Duffing system (33) show complex transition characteristics, such as single-well, double-well and triple-well characteristics analysis.

For the generalized fifth-order Duffing type system (33) when $A_5 > 0$, the bifurcated parameter sets $B_{B1} = B_{B1} \cup B_{H1} \cup B_{D1}$ in $(A_1, A_3, A_5)$ parametric space of bifurcation set, doubled set and double limit set as following

$$\begin{cases} B_{B1} = \{(A_1, A_3, A_5) \mid A_1 = 0, A_3 < 0, A_5 > 0\} \\ B_{H1} = \{(A_1, A_3, A_5) \mid A_1 = 0, A_3 > 0, A_5 > 0\} \\ B_{D1} = \{(A_1, A_3, A_5) \mid A_1 > 0, A_3 < 0, A_5 > 0, A_3^2 - 4A_1A_5 = 0\} \end{cases} \tag{34}$$

As shown in Fig. 11(a), the bifurcation set surface $B_1$ surface of the system (33) are plotted in three dimensional parametric $(A_1, A_3, A_5)$ space. As illustrated in Fig. 11(a), the bifurcation surface branches $B_{B1}$ is the super pitchfork bifurcation. $B_{H1}$ is the sub pitchfork bifurcation, saddle-node bifurcation surface $B_{D1}$. The space I, II, III corresponding the monostable, bistable and tristable well state respectively, for detail of phase portraits see Figs. 10(a-f).

For $A_5 < 0$, the bifurcated parameter sets $B_{B2} = B_{B2} \cup B_{H2} \cup B_{D2}$ in $(A_1, A_3, A_5)$ parametric space of bifurcation set, doubled set and double limit set as following

$$\begin{cases} B_{B2} = \{(A_1, A_3, A_5) \mid A_1 = 0, A_3 < 0, A_5 < 0\} \\ B_{H2} = \{(A_1, A_3, A_5) \mid A_1 = 0, A_3 > 0, A_5 < 0\} \\ B_{D2} = \{(A_1, A_3, A_5) \mid A_1 < 0, A_3 > 0, A_5 < 0, A_3^2 - 4A_1A_5 = 0\} \end{cases} \tag{35}$$

Fig. 11(b) Bifurcation surface branches $B_{B2}$ is the super pitchfork bifurcation. $B_{H2}$ is the sub pitchfork bifurcation, saddle-node bifurcation surface $B_{D2}$. The space regions IV, V, VI corresponding the mono-stable, bi-stable and tri-stable well characteristic, respectively.

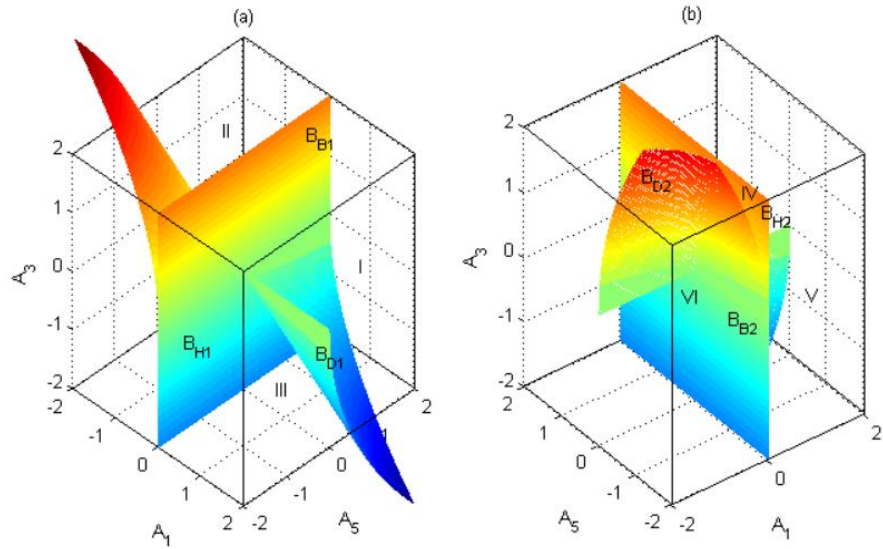

Fig. 11 Bifurcation sets of the nonlinear system (33). (a)$A_5 > 0$, (b) $A_5 < 0$ (Color online).

The phase portraits of nonlinear system (33) corresponding to the space regions of Fig. 11(b) are presented in Fig. 12. As shown in Fig. 12(a) phase portrait with the heteroclinic orbits on is drawn corresponding to bifurcation set $B_{B2}$ and the region IV. The phase portraits with a hyperbolic saddle point of Fig. 12(b) corresponding to the region V and the double limit bifurcation set $B_{D2}$. In Fig. 12(c), the phase portrait with a hyperbolic saddle point is plotted corresponding the hysteresis bifurcation set $B_{H2}$. As illustrated in Fig. 12(d), the phase portrait with homoclinic trajectories and double potential well is plotted corresponding region VI.

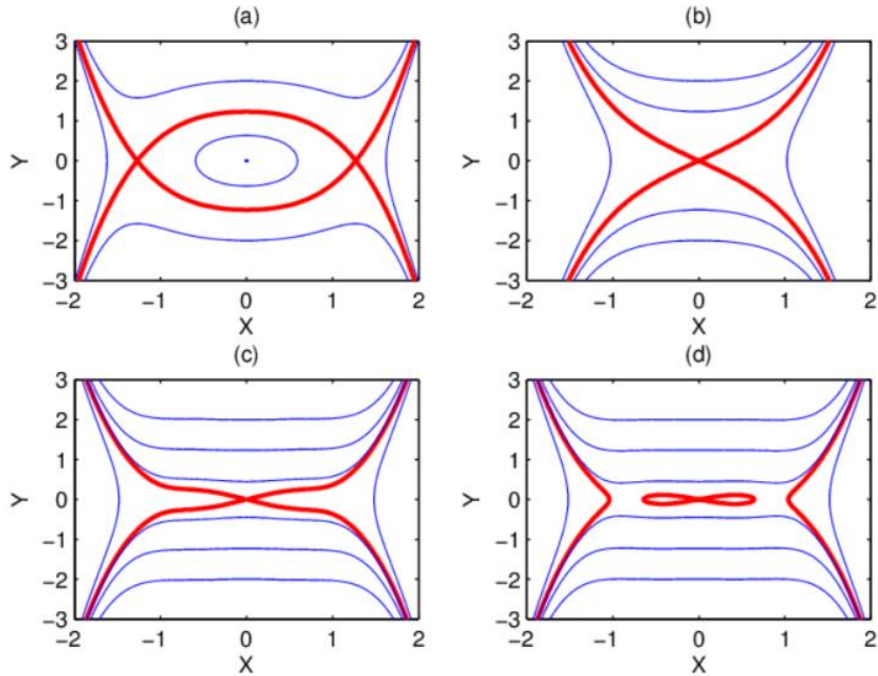

Fig. 12 Phase portraits of system (33) with help of Hamilton $H = KE + PE$. (a)At $O$, region IV and set $B_{B2}$, (b)on set $B_{D2}$ and region V, (c)on set $B_{H2}$, (d)and region VI (Color online).

### 4.2 Codimension bifurcation analysis of multiple limit cycles

For nonlinear damping force disappear $F_d \neq 0$, then the dynamic system of mechanical part of FIV energy harvesting system (16) is

$$X'' + F_d + F_s = 0 \tag{36}$$

The nonlinear friction damping force $F_d$ is approximated by

$$F_d = -\xi X' + \eta X'^3 \tag{37}$$

In order to predict the appearance of limit cycle and the instability characteristic, the codimension two bifurcation with parameters setting of $\mu = 0$, the mechanical system (36) with self-excited vibration and van der Pol type damping can be obtained as follows

$$X'' - \xi X' + \eta X'^3 + A_1 X + A_3 X^3 + A_5 X^5 = 0 \tag{38}$$

The codimension bifurcation set $B_3 = B_h \cup B_{sc} \cup B_{po}$ on $(A_1, \xi)$ plane for the self-excited vibration system (38) is defined as follow [26]:

$$\begin{cases} B_h = \{(A_1, \xi) \mid A_1 - \xi = 0\} \\ B_{sc} = \{(A_1, \xi) \mid 0.8 A_1 - \xi = 0\} \\ B_{po} = \{(A_1, \xi) \mid 0.752 A_1 - \xi = 0\} \end{cases} \tag{39}$$

where $B_h$ is the Hopf bifurcation set. $B_{sc}$ is the homoclinic saddle connection bifurcation set. $B_{po}$ is the periodic orbit bifurcation set. As seen in Fig. 13(a), the bifurcation set $B_3$ divide the parameter plane $(A_1, \xi)$ into four regions marked as I, II, III and IV respectively.

When the codimension bifurcation set $B_3$ is projected on the parametric $(\beta, \xi)$ plane, the bifurcation set $B_4 = B_h \cup B_{sc} \cup B_{po}$ is given by the following

$$\begin{cases} B_h = \{(\beta, \xi) \mid A_1(\beta) - \xi = 0\} \\ B_{sc} = \{(\beta, \xi) \mid 0.8 A_1(\beta) - \xi = 0\} \\ B_{po} = \{(\beta, \xi) \mid 0.752 A_1(\beta) - \xi = 0\} \end{cases} \tag{40}$$

where $B_h$ is the Hopf bifurcation set, $B_{sc}$ is the homoclinic saddle connection bifurcation set, $B_{po}$ is the periodic orbit bifurcation set. The codimensional two bifurcation set are plotted in Fig. 13(b) for the given parameters $\alpha = 0$, $\eta = 1$, $A_1 = -1$, $A_3 = 0$, $A_5 = 0$. the bifurcation set $B_4$ divide the parameter plane $(\beta, \xi)$ into four regions marked as I, II, III and IV respectively.

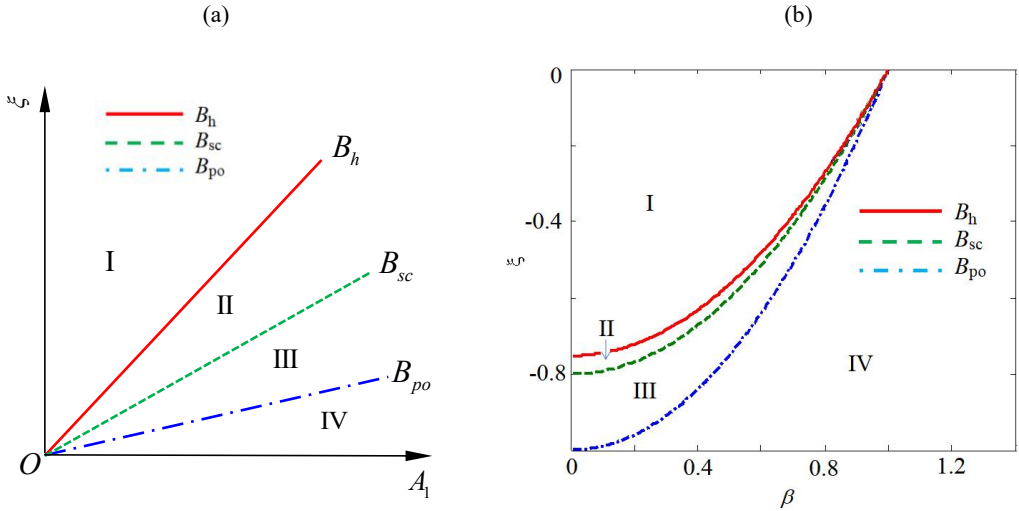

Fig. 13 Co-dimension bifurcation two [26] for bi-stable system (38). (a)On $(A_1, \xi)$ plane. (b) In $(\beta, \xi)$ plane (Color online).

In Fig. 14, the phase portraits of system (38) for single well case are plotted. Fig. 14(a) shows one closed orbit exist and its encircling all three fixed point. Fig. 14(b) depicted coexisting of one closed stable orbit and one inner unstable orbit. Fig. 14(c) presents the coexisting of one closed stable orbit and hetero-clinic saddle-center connection orbit. Fig. 14(d) gives coexisting of one closed stable orbit and outer unstable orbit. Fig. 14(e) illustrated the periodic orbit coalesce one unstable orbit. As illustrated in Fig. 14(f) the unstable orbit vanish.

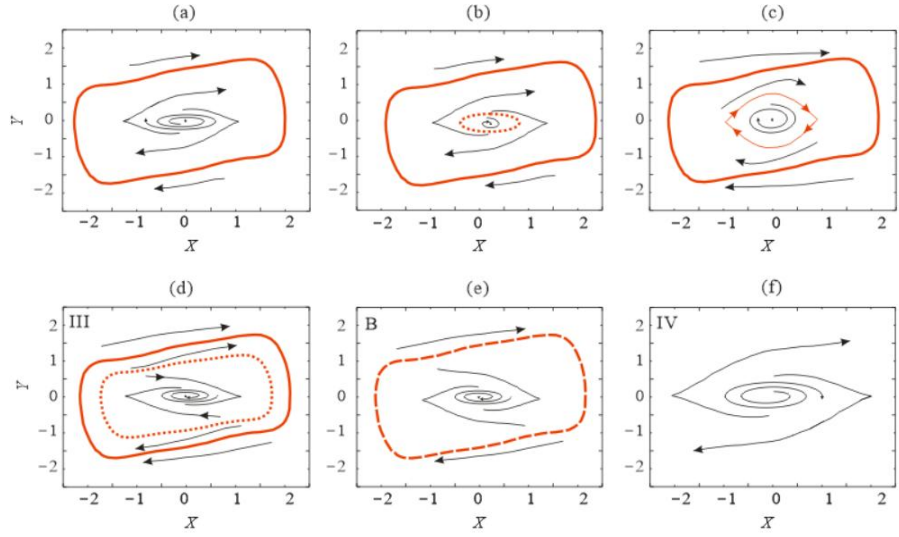

Fig. 14 Phase portraits of limit cycles for system (38) with SW. (a)In the region I and on Hopf bifurcation set $B_h$. (b)In rigion II. (c)On saddle connection bifurcation set $B_{sc}$. (d)In region III. (e)On periodic orbit bifurcation set $B_{po}$. (f)In region IV (Color online).

As shown in Fig. 15, the phase portraits of system (38) for double well case are illustrated. Fig. 15(a) shows one closed orbit exist and its encircling all three fixed point. Fig. 15(b) depicts coexisting of one closed stable orbit and one inner unstable orbit. Fig. 15(c) presents coexisting of one closed stable orbit and homo-clinic saddle connection orbit. Fig. 15(d) gives coexisting of one closed stable orbit and outer unstable orbit. orbit. Fig. 15(e) illustrated the periodic orbit coalesce one unstable orbit. Fig. 15(f) presents the unstable orbit vanish.

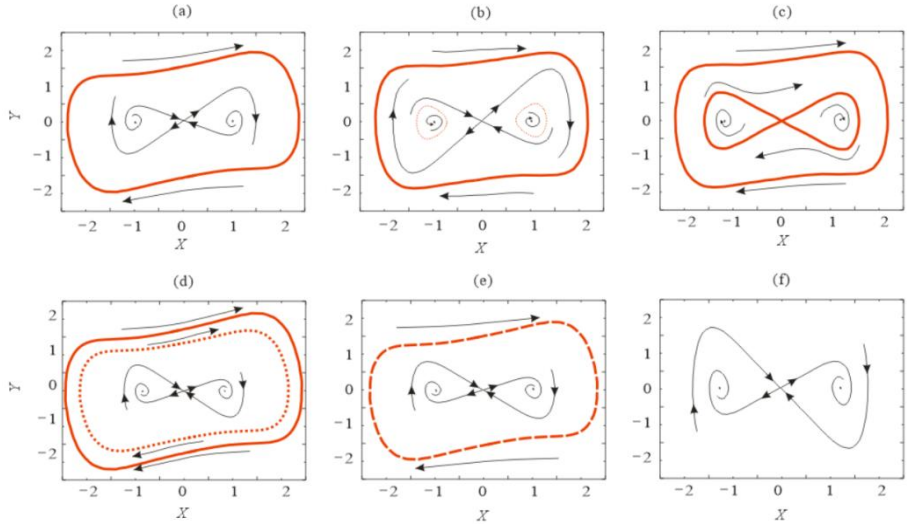

Fig. 15 Phase portraits of limit cycles for BS system (38) with DW. (a)In the region I and on Hopf bifurcation set $B_h$. (b)In rigion II. (c)On saddle connection bifurcation set $B_{sc}$. (d)In region III. (e)On periodic orbit bifurcation set $B_{po}$. (f)In region IV (Color online).

In Fig. 16, the phase portraits of system (38) with triple well are presented. Fig. 16(a) shows one closed orbit exist and its encircling all three fixed point. Fig. 16(b) depicted coexisting of one closed stable orbit and one inner unstable orbit. Fig. 16(c) presented coexisting of one closed stable orbit and hetero-homo-clinic saddle connection orbit. Fig. 16(d) give coexisting of one closed stable orbit and outer unstable orbit. Fig. 16(e) illustrated the periodic orbit coalesce one unstable orbit. Fig. 16(f) illustrated the unstable orbit vanish.

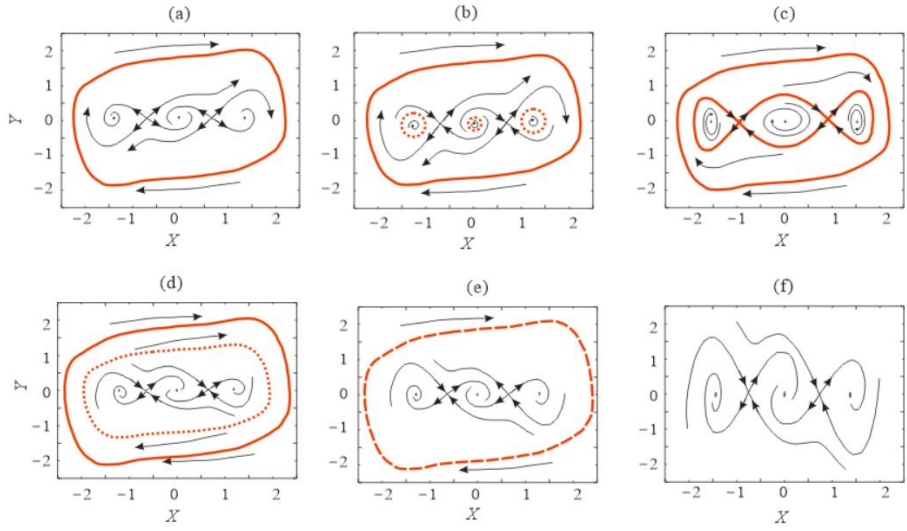

Fig. 16 Phase portraits of limit cycles for the TS system (38) with TW. (a)In the region I and on Hopf bifurcation set $B_h$. (b)In rigion II. (c)On saddle connection bifurcation set $B_{sc}$. (d)In region III. (e)On periodic orbit bifurcation set $B_{po}$. (f)In region IV (Color online).

4.3 Stick-slip motion with multiple limit cycles

For parameters $\mu_s = 0.1$, $v_m = 0.1$, $\xi = -0.8$, $\eta = 0.1$, $\gamma = 1.0$, the phase portraits of stick-slip motion for the friction-introduced system (36) are represented in Fig. 17. Based on the value of relative sliding velocity $V_r$, the system (36) exhibits two types of motion stage: kinetic slip stage for $V_r \neq 0$ referred to as dynamic and static stick stage for $V_r = 0$ referred to as breakaway.

Figs. 17(a, b) shown trajectory and limit cycle for SW system with mono-stable state for parameters setting of $\alpha = 0$, $\beta = 1.0$. We note that how initially vibration small spiral out to the limit cycle. With the setting of parameters $\alpha = 0.25$, $\beta = 2.5$, Figs. 17(c, d) depicted trajectories and coexistence of two limit cycles for DW system with bi-stable state. Figs. 17(e, f) presented trajectories and coexistence of three limit cycle for TW system with tri-stable state by using parameters of $\alpha = 0.5$, $\beta = 0.2$.

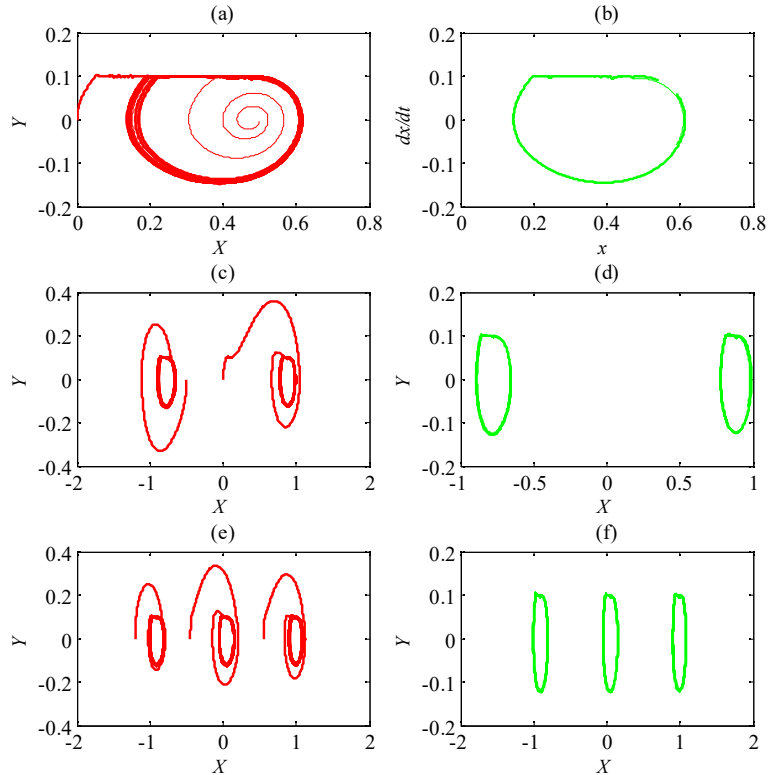

Fig. 17 Phase portraits of stick-slip motion for system (36) with monostable, bistable and tristable energy potential. (a) and (b) are the trajectories and limit cycles for SW system. (c) and (d) are the trajectories and limit cycles for DW system. (e) and (f) are the trajectories and limit cycles for TW system (Color online).

# 5. Influence of parameters on harvested electrical power output

## 5.1 Amplitude response analysis

The polynomial form of approximation of FIV energy system (16) are obtained as follows

$$\begin{cases} X'' - \xi X' + \eta X'^3 - \theta Q' + A_1 X + A_3 X^3 + A_5 X^5 = 0 \\ \gamma Q'' + \xi_q Q' - \theta X' + Q = 0 \end{cases} \quad (41)$$

Letting harmonic solution in complex form as

$$\begin{pmatrix} X \\ Q \end{pmatrix} = \begin{pmatrix} A_X \\ A_Q \end{pmatrix} e^{i\Omega T} \quad (42)$$

Where $i = \sqrt{-1}$ is the imaginary unit.

Submitting the harmonic solution (43) into the Eq.(43)

$$\left[ -\Omega^2 \begin{pmatrix} 1 & 0 \\ 0 & \gamma \end{pmatrix} + \Omega i \begin{pmatrix} -\xi + \eta A_1^2 & -\theta \\ -\theta & \gamma \end{pmatrix} + \begin{pmatrix} A_1 + A_3 X_X^2 + A_5 X_X^4 & 0 \\ 0 & 1 \end{pmatrix} \right] \begin{pmatrix} A_X \\ A_Q \end{pmatrix} e^{i\Omega T} = 0 \quad (43)$$

When $A_X$ and $A_Q$ are not equal to zero,

$$\begin{vmatrix} -\Omega^2 + \Omega i (-\xi + \eta A_1^2) + A_1 A_X + A_3 X_X^2 + A_5 X_X^4 & -\Omega \theta i \\ -\Omega \theta i & -\Omega^2 \gamma + 1 + \xi_q i \end{vmatrix} = 0 \quad (44)$$

Eq. (44) can be simplified in the following form

$$\begin{vmatrix} a_1 + a_2 i & a_3 i \\ a_3 i & a_4 + a_5 i \end{vmatrix} = 0 \quad (45)$$

where the coefficients in Eq. (45) are

$$\begin{cases} a_1 = A_1 + A_3 A_X^2 + A_5 A_X^4 - \Omega^2, \\ a_2 = \Omega(\xi - \eta A_X^2), \\ a_3 = -\Omega \theta, \\ a_4 = -\Omega^2 \gamma + 1, \\ a_5 = \Omega. \end{cases} \quad (46)$$

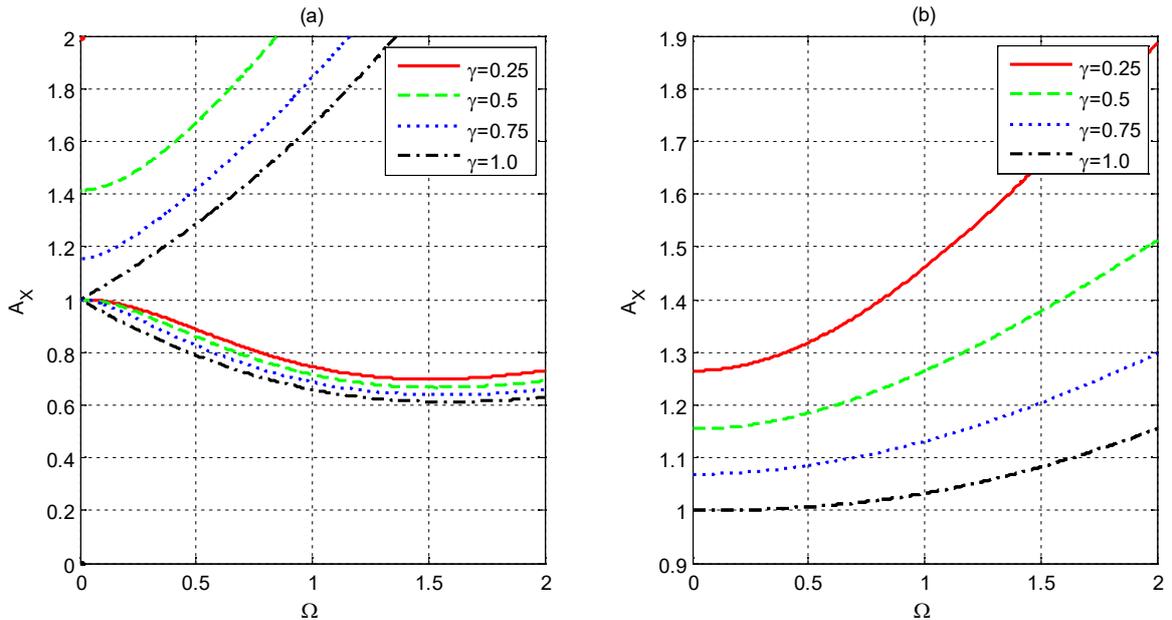

Fig. 18 Amplitude frequency response of FIV energy harvesting system (41) under different inertia ratio $\gamma$.
(a) For Eq. (47). (b) For Eq. (48) (Color online).

From the reality part of the Eq. (45), the nonlinear response relationship between amplitude, frequency and exciting damping coefficient becomes

$$\{(A_X,\Omega)\,|\,a_1a_4 - a_2a_5 + a_3^2 = 0\} \tag{47}$$

And from the imaginary part, the amplitude frequency relationship is that

$$\{(A_X,\Omega)\,|\,a_1a_5 + a_2a_4 = 0\} \tag{48}$$

By amplitude frequency relationship of formula of Eqs. (47) and (48), Figs. 18-21 give the nonlinear response curves of amplitude frequency of FIV system (41).

The response amplitudes $A_X$ as function of vibration frequency $\Omega$ of Eqs. (47) and (48) are are plotted in Fig. 18. As depicted in Fig. 18(a), it is found that the amplitude response exhibits the mutiple values characteristic for fixed frequency $\Omega$. When the dimensionless frequency $\Omega$ increasing, the amplitudes $A_X$ of the mechanical system rise of the up branch and first decrease and then riser of the down branch. It is observed that when the inertia ratio $\gamma = 0.25, 0.5.4, 0.75$ and $1.0$, the amplitude response curves become smaller for the bigger he inertia ratio. In Fig. 18(b), the response amplitude $A_X$ monotone increasing with frequency $\Omega$ increasing. It is also observed that amplitude become smaller with increasing of parameter $\gamma$ inertia ratio. Therefore, the increasing the inertia ratio could rise the output power from the point of view of energy convention.

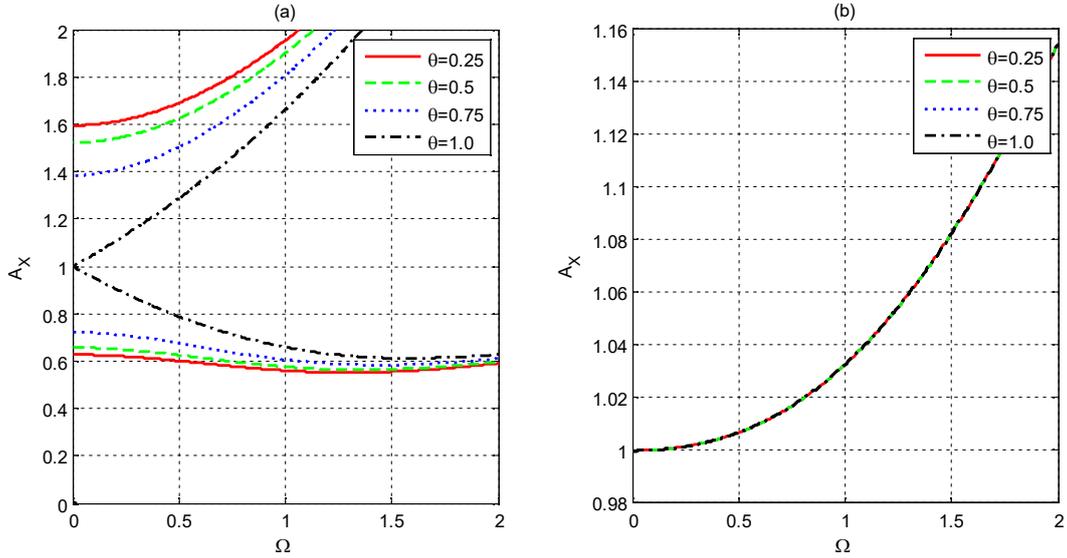

Fig. 19 Amplitude frequency response of FIV energy harvesting system (41) with different coupling coefficient ratio $\theta$.
(a)For Eq. (47). (b) For Eq. (48) (Color online).

Fig. 19 shows the effect of coupling ratio $\theta$ on the amplitude frequency curves of FIV energy harvesting system (41). In Fig. 19(a), it is noted that the amplitude response also behaviors the multiple values characteristic and variation tendency comparing to Fig. 18 (a). Interestingly, the amplitude response of Eq. (48) merge into one curve for different coupling ratio $\theta = 0.25, 0.5.4, 0.75$ and $1.0$. That is to say, the parameter have non influence on the response amplitude, which is not according with engineering practice and further discussion is need.

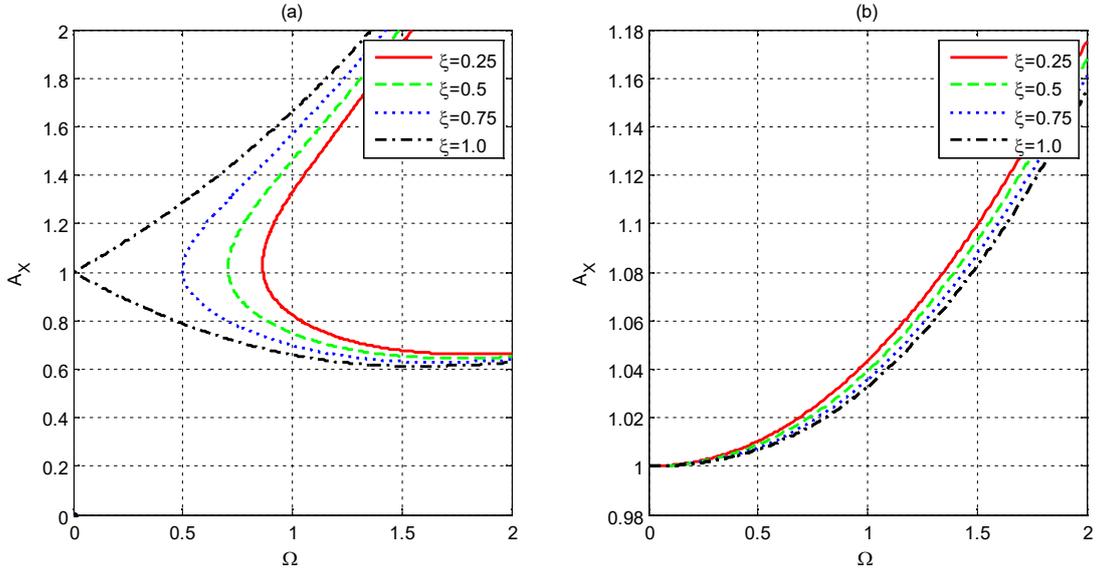

Fig. 20 Amplitude frequency response of FIV energy harvesting system (41) with different friction coefficient $\xi$.
(a) For Eq. (47). (b) For Eq. (48) (Color online).

As illustrated in Fig. 20, the impact of dimensionless damping ratio $\xi$ on the response amplitudes $A_X$ as function of vibration frequency $\Omega$ of Eqs. (47) and (48) are are plotted. In Fig. 20(a), when the damping ratio increases, the response amplitudes are vanished for the smaller vibration frequency $\Omega$. In the physical point of view, the enough bigger damping leading to the dead of vibration. As depicted in Fig. 20(b), we plot he response amplitude increase depending on the frequency $\Omega$ with the different value of friction ratio $\xi$. It is found that the increase the friction ratio lead to the decrease response amplitude, which mean that its benefit to the power harvesting.

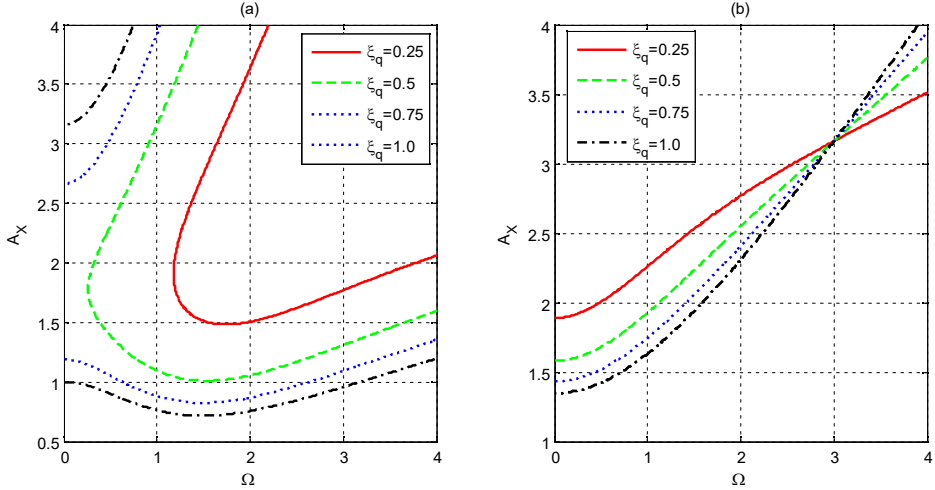

Fig. 21 Amplitude frequency response of FIV energy harvesting system (41) with different electrical resistance ratio $\xi_q$.
(a) For Eq. (47). (b) For Eq. (48) (Color online).

The response amplitudes $A_X$ as function of vibration frequency $\Omega$ of Eqs. (47) and (48) for different values of electrical resistance ratio $\xi_q$ are are plotted in Fig. 21. In Fig. 21(a) shows with an increase in frequency, the upper response amplitude branch increase and lower branch first decrease and then increase. It easy to find that the response exhibits multiple values phenomena. In Fig 21(b), it is seen from this figure that the amplitude curves cross in a fixed point around frequency $\Omega = 3$. When the frequency less than $\Omega = 3$, the bigger electrical resistance ratio, the smaller amplitude. On the contrary, while frequency more than $\Omega = 3$, the resistance coefficient

Hence, two analytical expression of the amplitude frequency Eqs. (47) and (48) help to understanding nonlinear mechanism of amplitude frequency response. However, this formula are remained open problem for the fact of the actual engineering vibration response appears to be unique, which is paradoxical multiple values of amplitude frequency response.

5.2 Response analysis

In this section, the parametric studies are carried out to explore the influence of the velocity, coupled, resistance and damping coefficients on the power output performance of the charge, current, voltage and power. For FIV energy harvesting system (16), the output of electronical voltage is

$$U = RI \tag{49}$$

The output of electronical power is

$$P = UI = RI^2 \tag{50}$$

In the numerical simulation, the values of the dimensional parameter are given in Table 2. The quasi zero stiffness of third order (QZS3) for $(\alpha_1, \beta_1) = (0, 1.0)$, the quasi zero stiffness of fifth order (QZS5) for $(\alpha_2, \beta_2) = (0.25, 0.72)$, bi-stable (BS) for $(\alpha_3, \beta_3) = (0.25, 0.5)$, triple stable (TS) for $(\alpha_4, \beta_4) = (0.5, 0.25)$.

Table 2. Non-dimensional parameters values used in the numerical simulation

| Parameter | Symbol | Value |
| --- | --- | --- |
| Horizontal geometrical ratio | $\alpha$ | $\alpha_i$, $i$=1, 2, 3, 4 |
| Vertical geometrical ratio | $\beta$ | $\beta_i$, $i$=1, 2, 3, 4 |
| Inertia ratio | $\gamma$ | 1.0 |
| Electra mechanical coupling ratio | $\theta$ | 0.1 |
| Mechanical damping ratio | $\xi_x$ | 0.1 |
| Electrical damping ratio | $\xi_q$ | 0.1 |
| Stribeck damping ratio | $\mu$ | 0.1 |
| Stribeck damping ratio | $\xi$ | 0.1 |
| Stribeck damping ratio | $\eta$ | 1.0 |

5.3 Effect of velocity

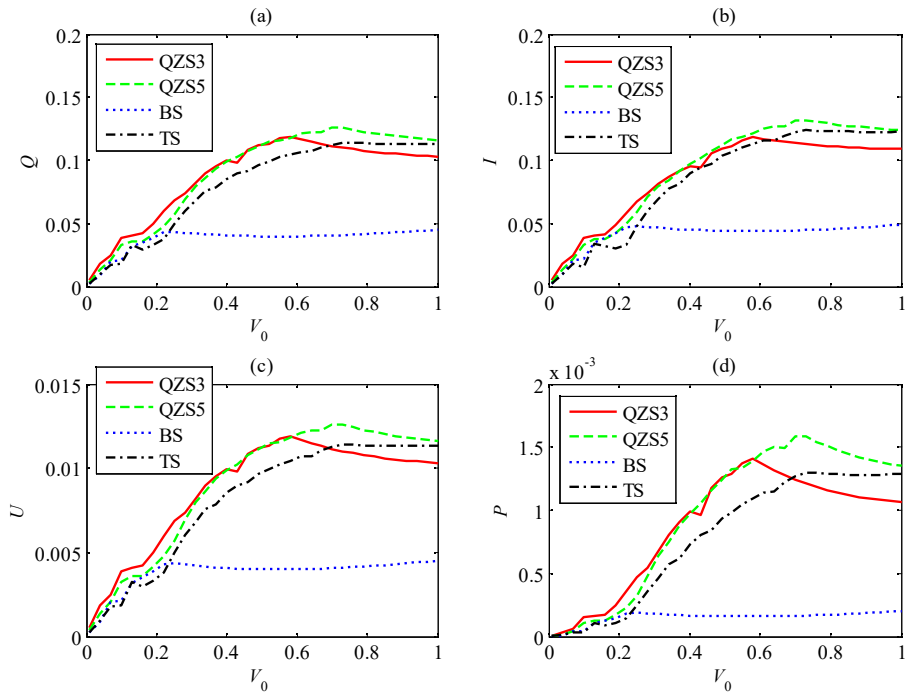

Fig. 22 The electrical output-velocity response of FIV system (16) for varying speed $V_0$. (a)Charge $Q$, (b) current $I$, (c)voltage $U$, (d)power $P$ (Color online).

The belt velocity $V_0$ was considered first. The qualitative aspects of the nonlinear electromagnetic coupled FIV system (16) can be obtained by numerical simulation method. The following system parameters and inertial conditions are chosen in Table. 1. A set of initial conditions $x(0) = 0$, $v(0) = 0$, $q(0) = 0$, $i(0) = 0$ are chosen. As shown in Fig. 22, the electrical outputs of charge and current voltage, power output for various value of $V_0$. The electrical output of $Q$, $I$, $U$ and $P$ for the QZS3, QZS5 and TS are the appearance of first increase and reach the critical high power, then deceasing. For the BS case, the output powers are increasing in the low frequency region and keep in the high frequency region. Therefore, the belt velocity is hardly impact on the output power of BS in the higher frequency region. To sum up, the QZS3 is the best appearance in the low frequency region and QZS5 is another best choice in the high frequency region.

### 5.4 Effect of electromechanical coupling efficient

The coupling coefficient $\theta$ is another parameter can influence the electrical power output. The coupling effect between the mechanical system and electrical system denoted by coupling coefficient $\theta$ at a range of [0, 1]. The qualitative aspects of the nonlinear electromagnetic coupling system (16) can be obtained by numerical simulation method. The following system parameters and inertial conditions are chosen in Table 2. A set of initial conditions $x(0)=0$, $v(0)=0$, $q(0)=0$, $i(0)=0$ are chosen. In Fig. 23, we show the electrical outputs of charge and current, voltage, power as a function of coupling efficient $\theta$ for multiple stability case of QZS3, QZS5, BS and TS respectively. As observe, the bigger that $\theta$ is, the larger is power output $Q$, $I$, $U$ and those dependence are linear of and $P$ is nonlinear relationship. On the whole, the electrical power increased as electromechanical coupled coefficient $\theta$ rise and the optimal parameter choices are QZS3 and QZS5.

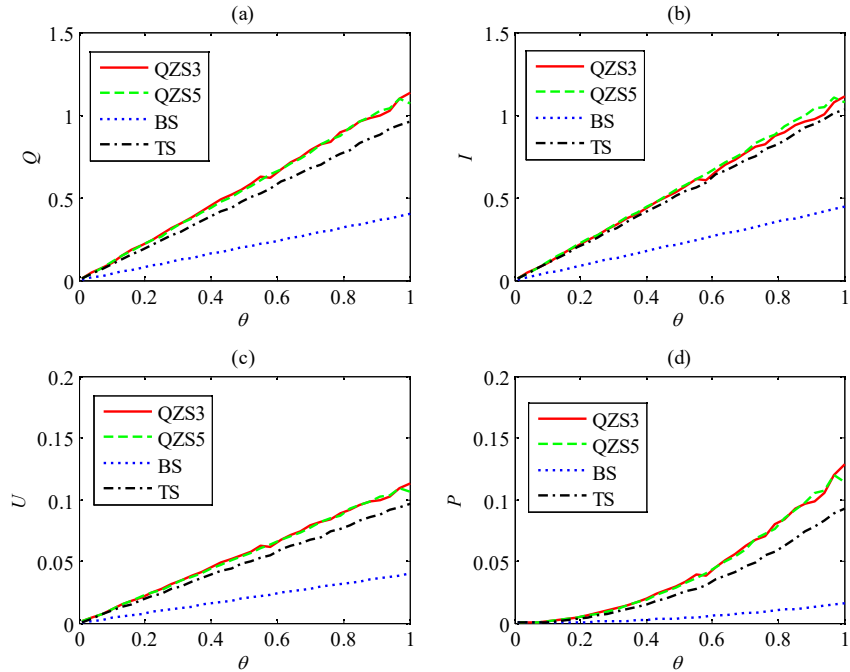

Fig. 23 The electrical output response of FIV system (16) for varying coupled parameter $\theta$. (a)Charge $Q$, (b) current $I$, (c)voltage $U$, (d)Power $P$ (Color online).

### 5.5 Effect of damping coefficient

The damping coefficient $\xi_x$ is more factor to controlling the electrical output power. The following parameters of system (16) and inertial conditions are chosen in Table 2. In Fig. 24, the electrical power outputs of $Q$, $U$, $I$ and $P$ are illustrated for the initial conditions $x(0)=0$, $v(0)=0$, $q(0)=0$, $i(0)=0$. Fig. 24 depicts the charge, voltage, current and power change versus damping coefficient $\xi_x$. It is found that the all electrical outputs are decease with increasing damping coefficient for all cases of QZS3, QZS5, BS and TS. It is found that the charges $Q$ is decease with increasing damping coefficient for all cases of QZS3, QZS5, BS and TS. In summary, it is clearly seen that the QZS3 is the optimal parameter arrangement to harvesting the environment vibration energy.

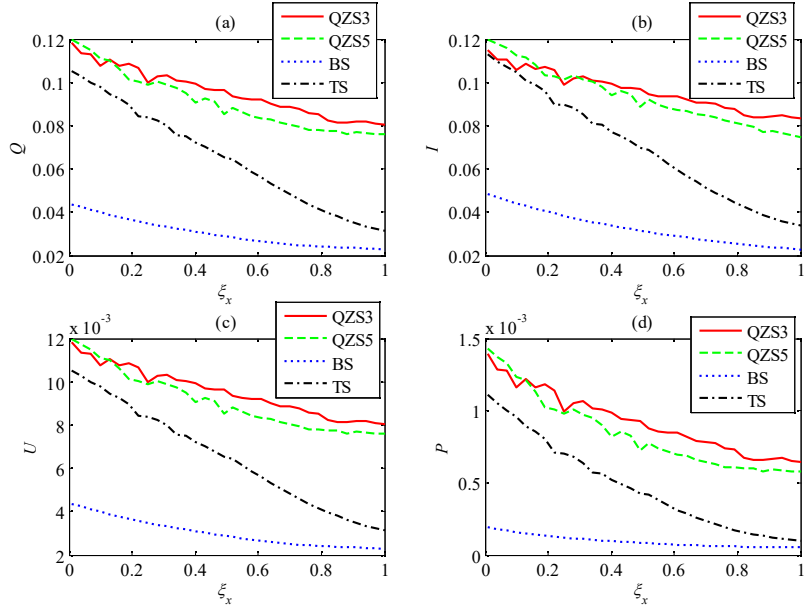

Fig. 24 The electrical output response of FIV (16) for different value of damping ratio $\xi_x$. (a)Charge $Q$, (b)current $I$, (c)voltage $U$, (d)Power $P$ (Color online).

5.6 Effect of resistance coefficient

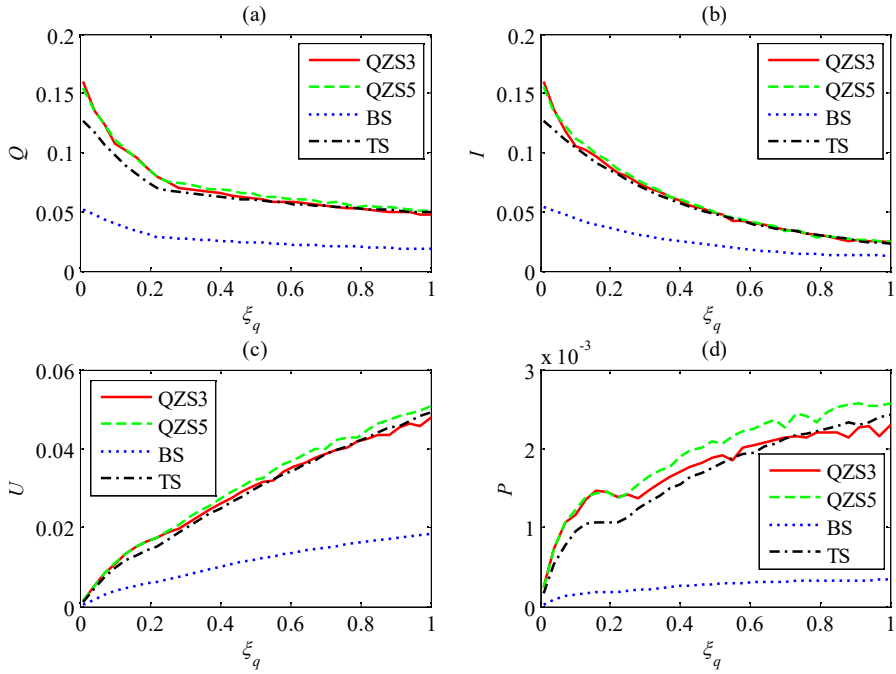

Fig. 25 The electrical output response amplitude of FIV system (16) for vary resistance parameter $\xi_q$. (a)Charge $Q$, (b) current $I$, (c)voltage $U$, (d)Power $P$ (Color online).

The electrical output of FIV system (16) is dependent on the external load resistance $\xi_q$ range of [0, 1] shown in Fig. 25. The following system parameters and inertial conditions are chosen in Table 2. A set of initial conditions $x(0)=0$, $v(0)=0$, $q(0)=0$, $i(0)=0$ are chosen. With the goal to optimize the performance of the energy conversion, we study how electrical outputs of charge and current voltage, power output depend on parameters $\xi_q$ of the energy harvester. It is fount that the charge $Q$ and current $I$ decreasing with increasing resistance ratio $\xi_q$. In the contrast, it can be noted that the voltage $U$ and electrical power $P$ decreasing with increasing resistance coefficient $\xi_q$. In particular, it should be noted that QZS5 is the highest harvested power in all frequency band of region [0, 1].

## 6. Experimental work

The experimental system is set up to verify the effectiveness of the sliding mode control algorithm of the proposed model.

6.1 The static force with multiple stability

In Fig. 26(a), the incline springs structure, ruler and dynamometer are designed to test the relationship of nonlinear force and displacement. The two linear springs are result to the nonlinear force because of the nonlinear geometrical configuration. The distance of rigids is $2a$, the height is $b$. the free length of spring is $l_0$, The ruler is used to tested the displacement deviation from the equilibrium point. The dynamometer is applied to measure the magnitude to restoring force in the horizontal direction pointing towards the equilibrium point.

In Fig. 26(b), we obtained static force of multiple well potentials to verify the QZS3, QZS5, DW and TW by the experimental device. It is shown that experimental results are aggree well with the theoretical curves of the QZS3, QZS5, DW and TW. The nondimensional parameters are $\alpha = a/l_0$ and $\beta = a/l_0$.

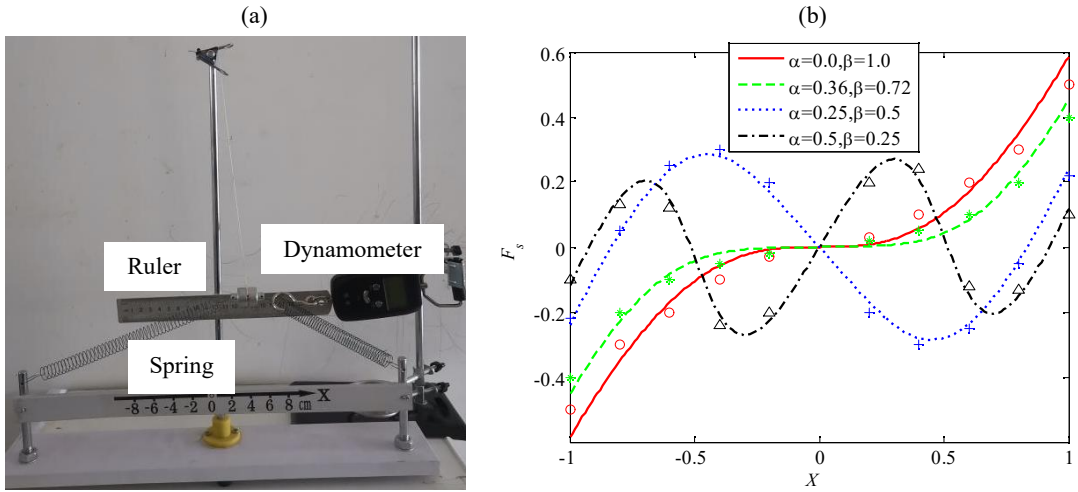

Fig. 26 Static force test. (a) Experiment setup, (b)Experimental force results. Symbol ○, *, + and △ denote the experimental data QZS3, QZS5, DW and TW respectively (Color online).

6.2 The nonlinear damping force

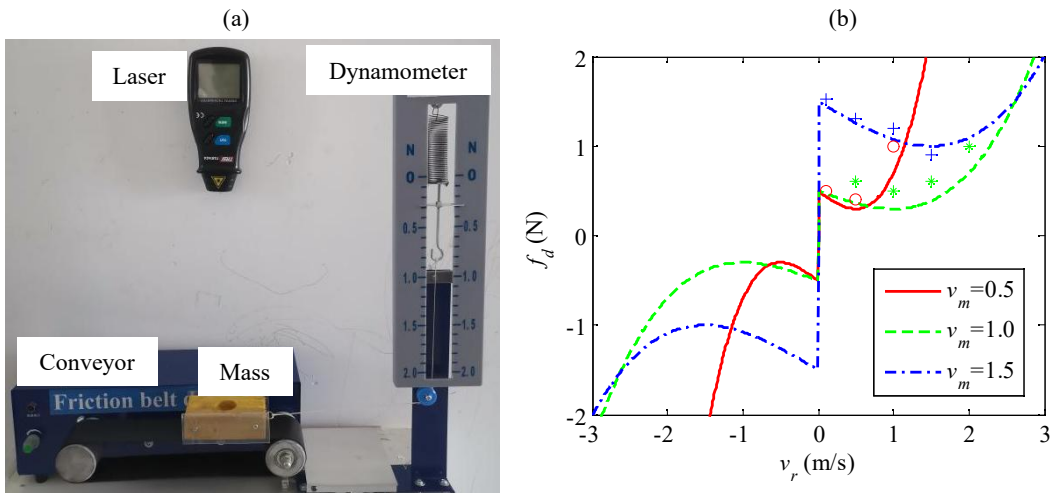

Fig. 27 Nonlinear damping forces. (a) Experiment setup, (b)Experimental force results. Symbol ○, * and + denotes the experimental data $v_m = 0.5$, $v_m = 1.0$ and $v_m = 1.5$ respectively (Color online).

As shown in Fig. 27(a), the conveyor, dynamometer and laser displacement sensor constructed to measure amplitude law of Stribeck friction. The dynamometer is used to determine the nonlinear friction force .The belt drive is a frictional drive that transmits power between two shafts using pulleys and an elastic belt. The laser sensor is applied to measure the displacement of the mass.

As presented in Fig. 27(b), the friction force velocity diagram are given to validate the theoretical results. The symbols ○, * and + denote the experimental data $v_m = 0.5$, $v_m = 1.0$ and $v_m = 1.5$ respectively. It is found that the experimental force values are consistent with the theoretical and numerical friction.

6.3 Experimental setup of the energy harvesting vibration

In illustrated in Fig. 28(a), a photograph of experiential rig for the energy harvesting device with QZS3, QZS5, DW and TW characteristic was demonstrated. The experimental devices consist of excited vibrator, mass-spring, electrical circuit, displacement laser. The displacement responses of mechanical vibration has been measured using of laser displacement sensor, which is used to measure the displacement of the mass *m*. The electrical response of electrical current *i* has been measured using measure device of multimeter. The belt conveying velocity is controlled by the frequency converter. The PC is used to captured steady-state response date of time history.

As shown in Fig. 28(b), the schematic of experiment setup is constructed to measure the displacement of mechanical system (MS) and current of electrical system (ES). As illustrated in Fig. 28(b), the devices of this experiment consist of MS excited by vibrator, ES, personal computer (PC) and displacement sensor (DS). Experimental study are carried out for the parameters in Table 3.

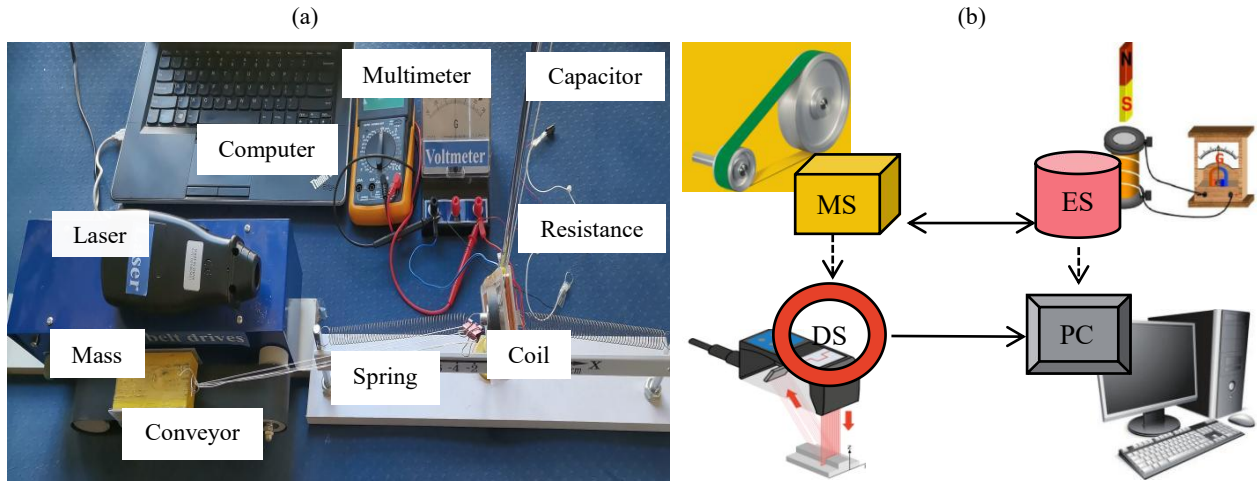

Fig. 28 Experiment setup. (a) Prototype FIV energy harvesting system, (b)Measuring devices schematic (Color online).

Table 3. Geometric and material properties used in experimental

| Parameter | Symbol | Value |
|---|---|---|
| Concentrated mass | $m$ | 0.2kg |
| Stiffness of oblique springs | $k$ | 1kN/m |
| Damping of damper | $c$ | 0.1N·s/m |
| Free length of oblique spring | $l_0$ | 0.2m |
| Length of electromagnetic coil | $l_e$ | 1m |
| Velocity of belt | $v_m$ | 0.3m/s |
| Static friction coefficient | $\mu_s$ | 0.06N·s/m |
| Dynamic friction coefficient | $\mu_m$ | 0.04N·s/m |
| Inductance of electromagnetic coil | $L$ | 1H |
| Capacitance of capacitor | $C$ | 1F |
| Resistance of resistor | $R$ | 1Ω |
| Magnetic flux density | $B$ | 0.1T |

From Fig. 29(a), it is shown that the simulation and experimental results for steady-state responses of voltage as the slider velocity $v_0$ increase. It is found that the the voltage of DW is decesing with the bigger belting velocity. In contrast, The output voltage of QZS3, QZS5 and TW are increasing with velocity.

As shown in Fig. 29(b), the response of the electrical power *P* are given. The symbols ○, *, + and △ denote the experimental data QZS3, QZS5, DW and TW respectively. It is evident that the harvester with BS can exhibits the highest harvesting power. To sum up, the experiment results are consistent with the numerical simulation response. The experimental verification of sliding mode control of variation of the output power for a mono-stable system is performed. The simulation results described in the above section show that stick-sliding mode motion can significantly improve the output power.

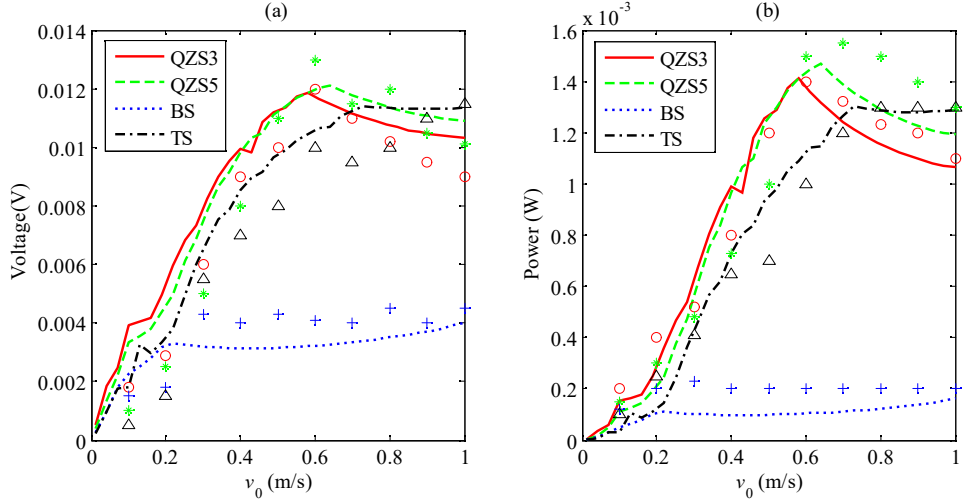

Fig. 29 Dynamical response of the experimental voltage and power of FIV energy harvesting system (10). Symbol ○, *, + and △ denotes the experimental data QZS3, QZS5, DW and TW respectively. (a)Output voltage. (b)Output power (Color online).

## 7. Conclusions

This paper established a FIV vibration energy harvesting system with both nonlinear elastic restoring force and nonlinear damping force due to the geometrical construction. For the mechanical part of FIV system, the restoring force of includes both irrational ($\sqrt{*}$) and quotient ($*/*$) non-linearity which behaviors multiple well potential and high-order quasi-zero stiffness dynamics. The damping force exhibits both quotient ($*/*$) and square ($*^2$) non-linearity. The Stribeck friction force, similar to the van der Pol, leads to the limit cycle of self-excited motion. The bifurcation sets of the equilibrium bifurcation surface, the codimension two and codimension three bifurcations and the corresponding phase portraits diagram are obtained to show the transition behavior of monostable, bistable and tristable stability and potential well. In addition, the analytical formula of amplitude frequency response are given to show the effects of the system parameters on the amplitude of displacement. The parametric dependencies of dimensionless velocity, coupling ratio, damping and resistance coefficients for FIV energy harvesting system with monostable QZS3 and QZS5, bistable DW and tristable TW, are numerically investigated to obtain the change trend of electrical power. At last, the experiment data of restoring force, nonlinear friction and power output are compared with theoretical and numerical results to verify proposed energy harvesting system.

We drew the conclusion like: (i) The analytical, numerical and experimental results of the nonlinear energy harvesting mechanism have motivated toward understanding of the stick-slip motion of friction induced vibration. (ii) The transition mechanism of the multiple well proposed here not limit the energy harvesting system, but also be applied to design the vibration isolation [31], vary stiffness robot [32], bionic mechanics [33], medical health [34]. (iii) The future works are focus on the the energy renewable technique of the machine health monitoring [35], the cutting tool [36] and treadmill[37].

**DRediT authorship contribution statement**

**Yanwei Han:** Conceptualization, Investigation, Writing-original draft. **Zijian Zhang:** Methodology, Funding acquisition, Writing review & editing.

**Declaration of Competing Interest**

The authors declare that they have no known competing financial interest or personal relationships that could have appeared to influence the work reported in this paper.

**Funding acknowledgement**

This work was supported by the State Key Laboratory of Robotics and System (HIT) (Grant no. SKLRS-2022-KF-19). Additionally, the authors would like to thank the anonymous reviewers and the editors for their relevant comments and useful suggestions.